\documentclass[referee]{raa}

\usepackage{graphicx,times}
\usepackage[numbers,sort&compress]{natbib}
\usepackage{amssymb,amsmath}
\bibpunct{[}{]}{,}{n}{}{,}
\let\cite\citep
\usepackage[scheme=plain]{ctex}
\usepackage{multirow}
\usepackage{setspace}
\usepackage{xcolor}
\newcommand{\githubchain}{\url{https://github.com/WhitaraAnderson/2607.21497}}
\newcommand{\githubaxiclass}{\url{https://github.com/PoulinV/AxiCLASS}}
\newcommand{\githubcamb}{\url{https://github.com/cmbant/CAMB}}
\newcommand{\githubcobaya}{\url{https://github.com/CobayaSampler/cobaya}}
\newcommand{\githubpolychord}{\url{https://github.com/PolyChord/PolyChordLite}}

\usepackage{url}
\usepackage{bm}
\usepackage{hyperref}
\hypersetup{
  colorlinks=true,
  linkcolor=cyan,
  urlcolor=blue,
  citecolor=red
}

\begin{document}

\title{Hubble tension: a short review of theoretical explanations}

\volnopage{26:084009 (25pp), 2026 August}
\setcounter{page}{1}

\author{Jia-Qi Wang\inst{1,2} \and Zong-Kuan Guo\inst{1,2,3}}

\institute{Institute of Theoretical Physics, Chinese Academy of Sciences (CAS), Beijing 100190, China; {\it wangjiaqi@itp.ac.cn, guozk@itp.ac.cn}\\
\and University of Chinese Academy of Sciences (UCAS), Beijing 100049, China\\
\and School of Fundamental Physics and Mathematical Sciences, Hangzhou Institute for Advanced Study, University of Chinese Academy of Sciences, Hangzhou 310024, China\\
\vs\no
{\small Received 2026 June 2; revised 2026 June 22; accepted 2026 June 25; published 2026 July 17}}

\abstract{The $\Lambda$ cold dark matter model successfully describes a wide range of cosmological observations. However, the persistent discrepancy between the value of the Hubble constant inferred from cosmic microwave background (CMB) measurements within this model and that obtained from local distance-ladder determinations points to a significant inconsistency. This short review examines theoretical responses across the cosmological inference chain, from the gravitational field equations to the pre-recombination sound horizon and the late-time distance-redshift relation. It discusses early-time and late-time mechanisms, as well as modified gravity, as ways to alter the acoustic ruler, distances, structure growth, or gravitational response. Current proposals can reduce the nominal tension, but often at the cost of correlated shifts in CMB spectra, standard-ruler distances, lensing, structure growth, or calibrator information. Further progress requires unified likelihoods and multi-probe tests linking all key observables under the same model assumptions. 
}

\authorrunning{J.-Q. Wang \& Z.-K. Guo}
\titlerunning{Hubble tension: a short review of theoretical explanations}
\date{}

\maketitle

\section{Introduction}\label{sec:intro}

The Hubble constant, $H_0$, measures the present expansion rate of the Universe
~\cite{Lemaitre:1927zz,Hubble:1929ig} and links local astrophysical distances
to the global cosmological model~\cite{HST:2000azd,Freedman:2010xv}.  In the
standard six-parameter \(\Lambda\) cold dark matter (\(\Lambda\)CDM) framework, whose empirical status rests on
the discovery of cosmic acceleration
~\cite{SupernovaSearchTeam:1998fmf,SupernovaCosmologyProject:1998vns} and on
the broader concordance picture~\cite{Peebles:2002gy,Weinberg:2013agg}, the
cosmic microwave background (CMB) does not measure $H_0$ directly.  The acoustic
pattern fixes the angular sound scale together with the physical
matter and baryon densities, using the well-understood physics of CMB
anisotropies~\cite{Hu:2001bc}.  This inference has reached sub-percent
precision with Planck~\cite{Planck:2018vyg}, after the earlier Wilkinson
Microwave Anisotropy Probe (WMAP) determination~\cite{WMAP:2012nax}, and is
now tested independently by high-resolution ground-based CMB measurements from
the South Pole Telescope third-generation survey (SPT-3G)
~\cite{SPT-3G:2022hvq} and the Atacama Cosmology Telescope (ACT)
~\cite{AtacamaCosmologyTelescope:2025blo}.

Low-redshift measurements reach the expansion scale through a different path.
They use calibrated distance ladders, geometric distances, time delays, standard
sirens, or velocity fields, with no reliance on the CMB acoustic ruler.  Many of
these methods prefer values of $H_0$ above the CMB-inferred value obtained in
$\Lambda$CDM, and this early--late discrepancy is usually called the Hubble
tension~\cite{Verde:2019ivm,DiValentino:2021izs,Abdalla:2022yfr}.  The issue
goes beyond the difference between two central values.  It compares two
inference chains, one anchored by pre-recombination standard-ruler physics and the
other by low-redshift distances and calibrations.

\begin{figure*}
\centering
\includegraphics[width=0.92\textwidth,height=0.68\textheight,keepaspectratio]{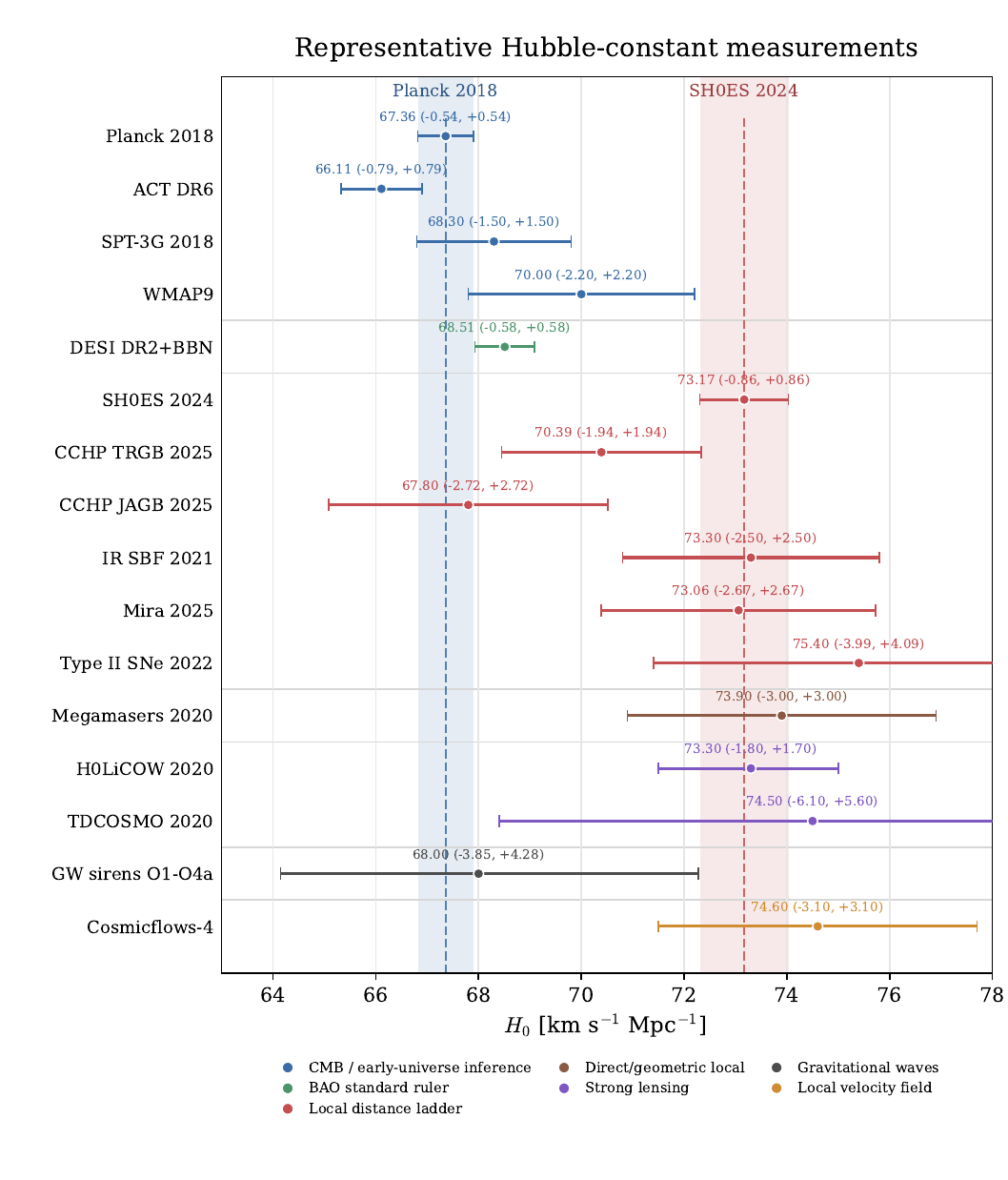}
\caption{
Representative recent determinations and inferences of $H_0$, compiled from
Refs.~\cite{WMAP:2012nax,Planck:2018vyg,SPT-3G:2022hvq,
AtacamaCosmologyTelescope:2025blo,DESI:2025zgx,Breuval:2024lsv,
Freedman:2024eph,Blakeslee:2021rqi,Bhardwaj:2025kbw,Pesce:2020xfe,
deJaeger:2022lit,H0LiCOW:2019pvv,Birrer:2020tax,Bom:2024afj,Tully:2022rbj}.
CMB-inferred values, standard-ruler measurements, and local or geometric
measurements have different model assumptions and statistical correlations.
}
\label{fig:h0_measurements}
\end{figure*}

Fig.~\ref{fig:h0_measurements} gives a compact overview of this landscape.
Its CMB-inferred entries show $\Lambda$CDM-dependent inferences from
WMAP9~\cite{WMAP:2012nax}, Planck 2018~\cite{Planck:2018vyg}, SPT-3G
~\cite{SPT-3G:2022hvq}, and ACT Data Release 6 (DR6)
~\cite{AtacamaCosmologyTelescope:2025blo}.
The standard-ruler entry uses Dark Energy Spectroscopic Instrument (DESI) Data
Release 2 (DR2) baryon acoustic oscillation (BAO) measurements with a
Big Bang nucleosynthesis (BBN) calibration~\cite{DESI:2025zgx}.  The local and
geometric side includes the Supernova \(H_0\) for the Equation of State (SH0ES)
Cepheid--Type Ia supernovae (SNe Ia) distance ladder~\cite{Riess:2021jrx,Breuval:2024lsv}, the Carnegie-Chicago Hubble Program (CCHP)
tip of the red giant branch (TRGB) and J-region asymptotic giant branch (JAGB) calibrations~\cite{Freedman:2024eph}, infrared
surface-brightness fluctuations~\cite{Blakeslee:2021rqi}, Mira variables
~\cite{Bhardwaj:2025kbw}, water megamasers~\cite{Pesce:2020xfe}, Type II
SNe~\cite{deJaeger:2022lit}, strong-lensing time delays from \(H_0\) Lenses in COSMOGRAIL's Wellspring (H0LiCOW)
~\cite{H0LiCOW:2019pvv} and TDCOSMO~\cite{Birrer:2020tax}, gravitational-wave
standard sirens~\cite{Bom:2024afj,Ruan:2019tje}, and the Cosmicflows-4 velocity-field
compilation~\cite{Tully:2022rbj}.  Recent community summaries of the local
distance network emphasize that these determinations share calibrators and
cross-checks, so the local side is best read as a calibration network rather
than a single datum~\cite{H0DN:2025lyy}.

Fig.~\ref{fig:h0_measurements} also shows why the issue has become a model-building problem.  CMB
and BAO analyses infer $H_0$ through the sound horizon as a standard ruler
~\cite{Eisenstein:1997ik,SDSS:2005xqv}, whereas the distance ladder and other
low-redshift methods infer it from nearby or intermediate-redshift distances
and velocities.  Within $\Lambda$CDM, Planck 2018 gives
$H_0=67.36\pm0.54\,{\rm km\,s^{-1}\,Mpc^{-1}}$~\cite{Planck:2018vyg}; the
SH0ES distance ladder gives a local value near
$73\,{\rm km\,s^{-1}\,Mpc^{-1}}$ with percent-level precision
~\cite{Riess:2021jrx,Breuval:2024lsv}.  Other late-time probes are generally
less precise or have different systematics, but several lie closer to the
local distance-ladder scale than to the Planck $\Lambda$CDM inference
~\cite{Blakeslee:2021rqi,Pesce:2020xfe,H0LiCOW:2019pvv,Tully:2022rbj}.  Thus
the tension is defined by the persistence of a
high local expansion scale when compared with the pre-recombination standard-ruler inference.

The statistical strength of this comparison depends on the adopted datasets and
diagnostic.
The SH0ES distance-ladder program includes the 2.4\% distance-ladder measurement
~\cite{Riess:2016jrr}, Gaia-based Milky Way Cepheid standards
~\cite{Riess:2018byc}, the Large Magellanic Cloud (LMC)-anchored 1\% calibration
~\cite{Riess:2019cxk}, the later comprehensive Hubble Space Telescope (HST) analysis
~\cite{Riess:2021jrx}, and high-precision cluster Cepheid calibration
~\cite{Riess:2022mme}, while James Webb Space Telescope (JWST) observations have directly tested Cepheid
crowding and related systematics~\cite{Riess:2023bfx,Riess:2024ohe}.  TRGB
calibrations give lower central values in some analyses
~\cite{Freedman:2019jwv,Freedman:2021ahq,Freedman:2024eph}, and flexible
time-delay lens modeling can broaden the allowed range~\cite{Birrer:2020tax}.
These qualifications matter for the quoted significance, but they have not
turned the discrepancy into a single-measurement anomaly.  Parameter-shift,
Bayesian evidence-ratio, and related concordance diagnostics
~\cite{Raveri:2018wln,Handley:2019wlz,Verde:2019ivm} show that the early--late
comparison is statistically significant under standard assumptions, while
studies of SN calibration and distance-ladder systematics
~\cite{Mortsell:2018mfj,Efstathiou:2020wxn,Efstathiou:2021ocp,Camarena:2021jlr} identify which
choices control the precise number of standard deviations.  This is the sense
in which the Hubble tension is treated here as a strong empirical challenge,
even though its numerical severity remains dataset-dependent
~\cite{DiValentino:2021izs,Abdalla:2022yfr}.

The theoretical difficulty is set by the sound horizon.  The CMB acoustic scale
is approximately
\begin{equation}
\theta_s=\frac{r_s(z_\ast)}{D_A(z_\ast)},
\end{equation}
where $r_s(z_\ast)$ is the comoving sound horizon at last scattering and
$D_A(z_\ast)$ is the angular-diameter distance to the last-scattering surface.  In
$\Lambda$CDM, raising $H_0$ while preserving the measured $\theta_s$ requires
correlated shifts in the physical matter density, the baryon density, and the
pre-recombination expansion history.  BAO data add a second standard-ruler
condition because low-redshift distances are measured relative to the
drag-epoch sound horizon $r_d$, the BAO ruler set at the baryon-drag epoch.
The Hubble tension is therefore often
rephrased as a sound-horizon problem
~\cite{Bernal:2016gxb,Aylor:2018drw,Knox:2019rjx}.  BBN-calibrated BAO analyses
provide a complementary standard-ruler route that does not require the full
Planck likelihood, while their precision also brings residual sensitivity to
light-element abundances, nuclear rates, and BBN codes
~\cite{Schoneberg:2022ggi,Schoneberg:2024ifp,Burns:2023sgx}.  These analyses show that
late-time-only models that raise $H_0$ without changing the ruler must still fit
CMB, BAO, and SN distances, while models that reduce the ruler must
preserve the CMB peak structure, damping tail, lensing, primordial
nucleosynthesis, and the growth of structure
~\cite{Schoneberg:2021qvd,Krishnan:2020obg,Jedamzik:2020zmd,Lin:2021sfs,Vagnozzi:2021gjh,Pedrotti:2024kpn}.

The DESI era has sharpened this standard-ruler test.  BAO evolved from the
first high-significance galaxy-clustering detection~\cite{SDSS:2005xqv} through
the completed Sloan Digital Sky Survey (SDSS) Baryon Oscillation Spectroscopic
Survey (BOSS) and extended BOSS (eBOSS) program~\cite{eBOSS:2020yzd} to the DESI measurements
~\cite{DESI:2024mwx,DESI:2025zgx}.  When DESI BAO are combined with CMB data
and SN samples such as Pantheon+~\cite{Scolnic:2021amr,Brout:2022vxf}
or Dark Energy Survey (DES) SNe~\cite{DES:2024jxu}, some combined
analyses find preferences for deviations from
\(\Lambda\)CDM.  DES weak-lensing and
galaxy-clustering analyses~\cite{DES:2021wwk}, ACT CMB spectra
~\cite{AtacamaCosmologyTelescope:2025blo}, and the broader pattern of
 cosmological tensions~\cite{Perivolaropoulos:2021jda,CosmoVerseNetwork:2025alb} have also motivated
renewed tests of $\Lambda$CDM.  These developments make late-time dynamics a
common target for parameterized tests, including $w_0w_a$CDM descriptions
based on the Chevallier--Polarski--Linder (CPL) form
~\cite{Chevallier:2000qy,Linder:2002et} and broader dark-energy model building
~\cite{Copeland:2006wr}.  In the context of the Hubble tension, their main role
is to sharpen late-time consistency tests.  Changing the low-redshift equation
of state can reshape distances, while the inferred impact on $H_0$ depends on
the data combination and parameterization unless the standard ruler is also
changed.  Recent DESI+SN discussions therefore motivate cautious wording: the
apparent preference for evolving dark energy can be shifted by supernova
calibration, sample choices, and low-redshift systematics, so it should not be
treated as a direct Hubble-tension solution by itself
~\cite{Efstathiou:2024xcq,Huang:2025som,DES:2025tir,Popovic:2025glk}.

Theoretical proposals can be organized by which part of the inference chain they
modify or reinterpret.  Early-time explanations modify the
pre-recombination expansion rate, the acoustic ruler, recombination, or
perturbation propagation.  Early dark energy (EDE) is the most developed example
~\cite{Karwal:2016vyq,Poulin:2018cxd,Kamionkowski:2022pkx,Poulin:2023lkg}, and related
possibilities include phase-transition EDE, dark radiation, self-interacting
neutrinos, and recombination-sector changes~\cite{Kreisch:2019yzn}.  Late-time
explanations change the low-redshift distance-redshift relation through
dynamical dark energy, vacuum-sector transitions, dark-sector energy transfer,
or local inhomogeneity and local-structure effects.  Modified gravity changes the gravitational
action, the effective Planck mass, or the relation between geometry, expansion,
lensing, and structure growth~\cite{Clifton:2011jh}.  In this organization, a
``non-minimal coupling'' is not a separate class of explanation.  An interaction
that triggers an early-time component is discussed with early-time physics; a
dark-matter--dark-energy exchange is part of late dark-sector dynamics; an
explicit $f(\phi)R$ term is treated as scalar-tensor modified gravity; and a
dark-sector coupling that changes the sector-dependent gravitational response is
discussed with gravity-related models.

This article focuses on theoretical explanations and uses observational
systematics mainly to define the empirical challenge.  For each proposal, we
ask which physical quantity is changed, how this affects the inferred
\(H_0\), and which CMB, BAO, SN, lensing, growth, local-gravity, or
stability tests restrict the change.  The timing of such a review is set by
new standard-ruler and CMB information: DESI and DES have sharpened the
low-redshift distance consistency problem, while ACT and SPT provide
independent high-resolution tests of pre-recombination physics.  Because EDE
remains the most developed early-time route, Sec.~\ref{sec:early} includes
reference Markov-chain Monte Carlo (MCMC) constraints for fixed-\(n\) and free-\(n\) axion-like EDE,
including DESI-era and DES SN combinations where appropriate
~\cite{DES:2025sig,Jhaveri:2026bla}.  These calculations are used as reference
constraints and are interpreted cautiously.  Sec.~\ref{sec:modified_gravity}
turns first to scalar-tensor gravity, non-minimally coupled quintessence, and
other modifications of the gravitational sector.  Sec.~\ref{sec:early} reviews
early-time explanations with EDE as the main representative model.
Sec.~\ref{sec:late} discusses late expansion, dark-sector energy transfer, and
local-inhomogeneity mechanisms.  The
final section compares what a successful theoretical explanation would need to
accomplish.

Throughout the theoretical discussion, scalar-field equations are written in
reduced Planck units, \(c=\hbar=M_{\rm P}=1\), with
\(M_{\rm P}=(8\pi G)^{-1/2}\), unless dimensionful factors are displayed
explicitly.

\section{Modified gravity}\label{sec:modified_gravity}

Modified gravity changes the gravitational part of the inference chain.  In this
section the term includes scalar-curvature theories such as \(f(\phi)R\),
evolving effective Planck mass models, curvature, torsion, and nonmetricity
actions, and infrared modifications with extra gravitational degrees of freedom.
It also includes non-minimally coupled quintessence models in which gravity is
standard for the visible sector but a scalar field changes the gravitational
response or effective dynamics of the dark sector.  These proposals are
different from the late-time energy-transfer models discussed in
Sec.~\ref{sec:late}.  They modify the gravitational action, the gravitational
field equations, or the sector-dependent gravitational response, so the
expansion history, lensing, matter growth, gravitational-wave propagation,
screening, and local tests are correlated.  Any attempt to raise the inferred
\(H_0\) in this class therefore has to pass tests of both the background and
perturbation sectors~\cite{Clifton:2011jh,Capozziello:2011et}.

\begin{figure*}
\centering
\includegraphics[width=0.92\textwidth,height=0.68\textheight,keepaspectratio]{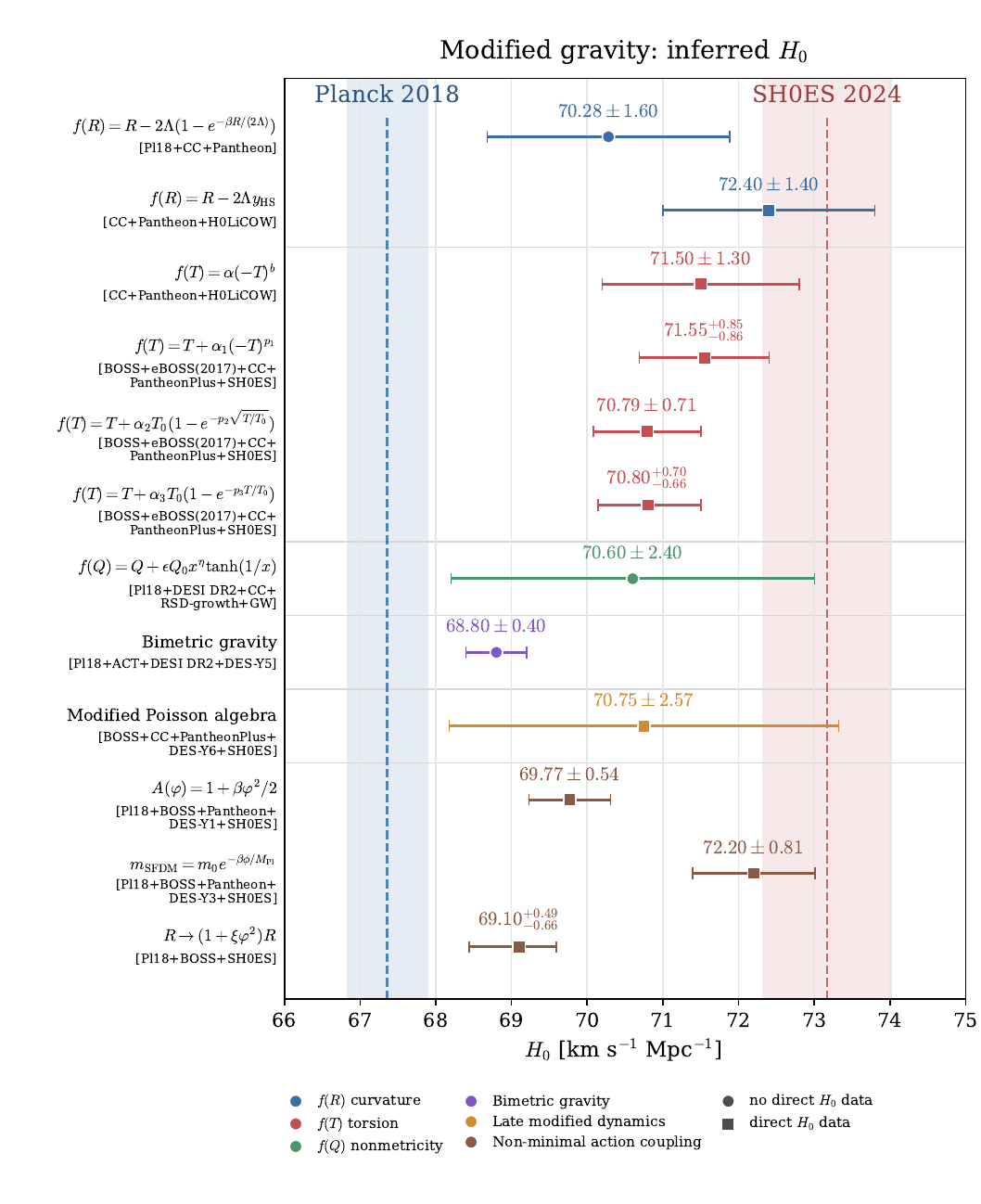}
\caption{
Representative \(H_0\) values in modified-gravity and gravity-related
non-minimal coupling proposals.  Circles denote analyses without direct
late-time \(H_0\) information in the quoted result, while squares denote data
combinations containing SH0ES, H0LiCOW, or an analogous direct late-time
calibration.  The vertical bands show the Planck 2018 and SH0ES 2024 reference
values.  The plotted entries are compiled from
Refs.~\cite{Odintsov:2020qzd,DAgostino:2020dhv,Briffa:2023ern,Kavya:2025vsj,
Hogas:2025ahb,Escamilla:2024xmz,Pitrou:2023swx,Uzan:2023dsk,Liu:2023kce,
Braglia:2020iik,Ye:2024zpk}.
}
\label{fig:modified_gravity_h0}
\end{figure*}

Fig.~\ref{fig:modified_gravity_h0} summarizes representative \(H_0\) values in
modified-gravity and gravity-related non-minimal action or coupling proposals.
The entries combine curvature, torsion, nonmetricity, infrared gravity,
scalar-tensor gravity, and species-dependent action couplings.  Some entries
include direct late-time information such as SH0ES or H0LiCOW, while others use
CMB, BAO, SNe, growth, or compressed distance information.  A high value of
\(H_0\) in Fig.~\ref{fig:modified_gravity_h0} therefore has to be interpreted
together with the physical modification, the perturbation sector, and the
external calibrator entering the fit.

\subsection{Scalar-tensor gravity}

Scalar-tensor gravity gives a representative example of how a modified
gravitational action can reproduce part of the pre-recombination expansion
change usually associated with EDE.  Thawing gravity provides a representative
example~\cite{Ye:2024zpk}.  Its cosmological effective action is
\begin{equation}
\begin{aligned}
S=&\int d^4x\sqrt{-g}\left[
\frac{1}{2}f(\phi)R+X-V(\phi)\right]\\
&+S_{\rm m}[g_{\mu\nu}],\\
X=&-\frac{1}{2}g^{\mu\nu}\nabla_\mu\phi\nabla_\nu\phi ,
\end{aligned}
\end{equation}
where \(S_{\rm m}\) is the matter action, \(R\) is the Ricci scalar, and
\(f(\phi)\) is the non-minimal coupling function.
Variation gives
\begin{equation}
\begin{aligned}
fG_{\mu\nu}+g_{\mu\nu}\Box f-\nabla_\mu\nabla_\nu f
&=T_{\mu\nu}^{(\phi)}+T_{\mu\nu}^{({\rm m})},\\
-\Box\phi&=\frac{1}{2}f_{,\phi}R-V_{,\phi},
\end{aligned}
\end{equation}
where \(T_{\mu\nu}^{(\phi)}\) and \(T_{\mu\nu}^{({\rm m})}\) are the scalar-field and
matter stress-energy tensors.  Commas denote derivatives with respect to
\(\phi\).
At fixed background curvature, the scalar equation takes the conventional
potential form \(-\Box\phi+V_{{\rm eff},\phi}=0\), with
\begin{equation}
V_{\rm eff}(\phi;R)=V(\phi)-\frac{1}{2}R f(\phi).
\end{equation}
For the choice \(V=V_0\exp(-\lambda\phi)\) and
\(f(\phi)=1-\xi\phi^2\), the spatially flat background equations become
\begin{equation}
3H^2\left(1-\xi\phi^2\right)-6\xi H\phi\dot{\phi}
=\frac{1}{2}\dot{\phi}^2+V+\rho_{\rm m},
\end{equation}
\begin{equation}
\ddot{\phi}+3H\dot{\phi}+V_{,\phi}
+6\left(2+\frac{\dot H}{H^2}\right)\xi H^2\phi=0 .
\end{equation}
Here \(\xi\) and \(\lambda\) are model parameters, \(\rho_{\rm m}\) is the matter
density, and overdots denote derivatives with respect to cosmic time.
During radiation domination, \(R\ll H^2\) and Hubble friction freezes the field.
Near matter-radiation equality, \(R\) becomes comparable to \(H^2\), the
curvature term in \(V_{\rm eff}\) becomes dynamically relevant, and the scalar
begins to roll.  The resulting change of \(M_{\rm eff}^2=f(\phi)\), or
equivalently of the pre-recombination Newtonian strength \(G_{\rm CMB}\), raises
the pre-recombination expansion rate and reduces the acoustic scale in a way
similar to EDE.  The same field also modifies the post-recombination
gravitational strength and can change the growth history.
In the fits of Ref.~\cite{Ye:2024zpk}, this combined pre- and
post-recombination gravitational response reduces the need for the large
\(\omega_{\rm cdm}\) shift that often appears in axion-like EDE, while keeping
\(S_8\) closer to weak-lensing values.  The theory must still be screened on
local scales and satisfy BBN, CMB lensing, LSS, and gravitational tests.
Earlier scalar-tensor effective-Planck-mass models use the same mechanism, for
example \(f(\phi)=(1+\xi \phi^n)\), where the non-minimal coupling changes the
expansion rate, BBN constraints, lensing, and growth together
~\cite{Braglia:2020iik}.

\subsection{Non-minimally coupled quintessence}

Non-minimally coupled quintessence can also affect the Hubble inference without
introducing a scalar-curvature term in the visible-sector gravitational action.
In \(f(\phi)R\) scalar-tensor gravity, the scalar changes the effective Planck
mass seen by all sectors.  Here the Standard-Model sector can remain minimally
coupled to \(g_{\mu\nu}\), while dark matter follows a different effective
metric,
\begin{equation}
S=S_{\rm grav}[g_{\mu\nu},\phi]+S_{\rm SM}[g_{\mu\nu},\psi_{\rm SM}]
+S_{\rm dm}[A^2(\phi)g_{\mu\nu},\psi_{\rm dm}],
\label{eq:dm_metric_coupling_v2}
\end{equation}
where \(A(\phi)\) controls the dark-sector coupling
~\cite{Pitrou:2023swx,Uzan:2023dsk}.  With
\(\alpha_A=d\ln A/d\phi\), pressureless dark matter obeys
\begin{equation}
\dot{\rho}_{\rm dm}+3H\rho_{\rm dm}
=\alpha_A(\phi)\rho_{\rm dm}\dot{\phi},\qquad
\rho_{\rm dm}\propto a^{-3}A(\phi),
\end{equation}
and the scalar equation receives a source proportional to
\(\alpha_A\rho_{\rm dm}\).  The coupling therefore changes the effective dark
matter density and gravitational response, rather than adding a separately
conserved EDE-like fluid.

The minimal realization of Refs.~\cite{Pitrou:2023swx,Uzan:2023dsk} sets
\(V(\phi)=0\) and takes \(A(\phi)=1+\beta\phi^2/2\).  The scalar energy density
remains subdominant and the scalar-mediated force acts only inside the dark
sector.  Visible matter, photons, and baryons still follow \(g_{\mu\nu}\), so
the standard acoustic physics and \(r_s\) are nearly unchanged.  The shift in
the \(H_0\) inference instead comes from the dark matter density sourcing the
background expansion, gravitational potentials, CMB lensing, and growth.  In
the fits of Ref.~\cite{Pitrou:2023swx}, CMB, BAO, and local-\(H_0\) data partly
lower the Planck--SH0ES discrepancy; removing the lowest-redshift BAO points
allows a larger \(H_0\) shift and reduces the tension to below about
\(2\sigma\).  This sensitivity to the BAO subset shows that the mechanism must
be tested with the full distance and perturbation data.  Since baryons and dark
matter do not respond to identical long-range forces, CMB spectra, lensing,
RSD, weak lensing, and dark-matter--baryon relative velocities provide important
consistency tests~\cite{Uzan:2023dsk}.

The related NMCQ model of Ref.~\cite{Wang:2026wrk} uses the same type of
dark-sector metric coupling but includes an exponential quintessence potential,
\begin{equation}
A(\phi)=1+\frac{1}{2}\beta\phi^2,\qquad
V(\phi)=V_0 e^{\alpha\phi}.
\end{equation}
The dark matter term modifies the scalar force and changes the dark-sector
energy balance, but its direct impact on the Hubble tension is limited in the
currently allowed parameter region.  Ref.~\cite{Wang:2026wrk} finds that the
coupling strength and the initial scalar amplitude have little direct effect on
the \(H_0\) posterior, mainly because a rapid post-equality change of the scalar
and dark matter sector is strongly constrained by CMB spectra and by the joint
distance data.  Related NMCQ implementations can be useful for DESI-era
dark-energy consistency tests~\cite{Wang:2025znm}, but they should be
distinguished from a direct Hubble-tension explanation.  These gravity-related
coupling entries are included in Fig.~\ref{fig:modified_gravity_h0}, whereas
Sec.~\ref{sec:late} discusses phenomenological energy-transfer models that
start from modified continuity equations.

\subsection{\texorpdfstring{\(f(R)\)}{f(R)} gravity}

Curvature-based \(f(R)\) gravity realizes the same scalar degree of freedom in a
pure curvature language.  The action is
\begin{equation}
S=\frac{1}{2}\int d^4x\sqrt{-g}\,f(R)+S_{\rm m},
\end{equation}
where \(S_{\rm m}\) is the matter action and \(f(R)\) replaces the
Einstein-Hilbert Lagrangian \(R\).
The metric field equations are
\begin{equation}
\begin{aligned}
f_R R_{\mu\nu}-\frac{1}{2}f g_{\mu\nu}
+\left(g_{\mu\nu}\Box-\nabla_\mu\nabla_\nu\right)f_R
&=T_{\mu\nu},\\
f_R &\equiv \frac{df(R)}{dR},
\end{aligned}
\end{equation}
where \(T_{\mu\nu}\) is the matter stress-energy tensor.
The scalaron \(f_R\) changes both the background expansion and the effective
force law.  If it is light on cosmological scales, it changes \(H(z)\) and
growth.  If it is screened in dense regions, the theory can approach general
relativity in the Solar System.  This scalaron picture is the essential physics
behind Hubble-tension applications of \(f(R)\) gravity
~\cite{Sotiriou:2008rp,DeFelice:2010aj,Nojiri:2010wj}.  A representative
late-time model used in Ref.~\cite{Odintsov:2020qzd} is
\begin{equation}
f(R)=R-2\Lambda\left[1-\exp\!\left(-\frac{\beta R}{2\Lambda}\right)\right],
\end{equation}
where \(\beta\) is a dimensionless model parameter and \(\Lambda\) sets the
late-time cosmological scale.
This model tends to \(\Lambda\)CDM for \(\beta\rightarrow\infty\) and at large
curvature.  It changes the late expansion by replacing a constant
\(\Lambda\)-term with a curvature-dependent contribution.  The same work also
studied a power-law model with an EDE-like curvature term,
\begin{equation}
\begin{aligned}
f(R)=&\,R-2\Lambda\gamma\left(\frac{R}{2\Lambda}\right)^\delta
      +F_{\mathrm{EDE}},\\
F_{\mathrm{EDE}}=&\,-2\alpha\Lambda\,
\frac{R^{m-n}(R-R_{\rm rec})^n}{R_{\rm rec}^{\ell+m}+R^{\ell+m}} .
\end{aligned}
\end{equation}
Here \(F_{\mathrm{EDE}}\) denotes the EDE-like curvature contribution to the
\(f(R)\) Lagrangian and is distinct from the fractional EDE density
\(f_{\rm EDE}\) used above.  The scale \(R_{\rm rec}\) is chosen near the
curvature scale relevant to the pre-recombination acoustic scale, while
\(\gamma\), \(\delta\), \(\alpha\), \(\ell\), \(m\), and \(n\) are model
constants.  The extra term behaves as a quasi-constant component near that epoch
and decays at late times, so it attempts to modify the same sound-horizon
physics as EDE inside a curvature action.  In practice, the inferred \(H_0\)
values in Ref.~\cite{Odintsov:2020qzd} remain intermediate and strongly
dataset-dependent, and the \(\Lambda\)CDM limit is often close to the allowed
region.  Related \(f(R)\) or Jordan-frame analyses
~\cite{Nojiri:2022ski,Schiavone:2022wvq,Montani:2023xpd,Bisabr:2024rfq,
Valletta:2025bgu,DOnofrio:2025cuk,Zhang:2026sxi,Chen:2024wqc} can be
understood as variations on this scalaron mechanism.  Time-delay-lens fits of
viable Hu--Sawicki-type \(f(R)\) models~\cite{Hu:2007nk,DAgostino:2020dhv}
further illustrate the direct-calibrator dependence.  They can accommodate high
\(H_0\) when H0LiCOW distances are included, while the modified-gravity
parameter remains statistically compatible with the GR limit.

\subsection{\texorpdfstring{\(f(T)\)}{f(T)} gravity}

Teleparallel gravity changes the geometrical variable from curvature to torsion.
For a flat FLRW metric the torsion scalar is \(T=-6H^2\), and the modified
action can be written as
\begin{equation}
S=\frac{1}{2}\int d^4x\,e\,[-T+f(T)]+S_{\rm m} ,
\end{equation}
where \(e\) is the tetrad determinant.
The background equation used in many \(f(T)\) analyses is
\begin{equation}
H^2+\frac{T}{3}f_T-\frac{f}{6}=\frac{1}{3}\rho ,
\end{equation}
where \(f_T=df(T)/dT\) and \(\rho\) is the total energy density.
Thus \(f(T)\) acts as an effective energy component whose value is tied directly
to \(H^2\).  The power-law representative
\begin{equation}
\begin{aligned}
f(T)=&\,\alpha(-T)^{p},\\
E^2(z)=&\,\Omega_{{\rm m}0}(1+z)^3+\Omega_{{\rm r}0}(1+z)^4
\\
&+\left(1-\Omega_{{\rm m}0}-\Omega_{{\rm r}0}\right)E^{2p}(z)
\end{aligned}
\end{equation}
is a useful example~\cite{Bengochea:2010sg,Briffa:2023ern}.  Here
\(E(z)=H(z)/H_0\), \(\Omega_{{\rm m}0}\) and \(\Omega_{{\rm r}0}\) are the
present-day matter and radiation density parameters, \(\alpha\) is fixed by the
present Friedmann equation, and \(p\) is the independent deformation parameter.
The parameter \(p\) changes the recent expansion rate through a nonlinear
function of \(H\).  In the PantheonPlus+SH0ES combinations studied in
Ref.~\cite{Briffa:2023ern}, the resulting \(H_0\) values move toward the local
calibration, while BAO helps break the degeneracy with \(\Omega_{{\rm m}0}\) and
\(p\).  Linder-type and exponential \(f(T)\) forms
~\cite{Hashim:2020sez,Hashim:2021pkq,Ren:2022aeo,Aljaf:2022fbk,
Bouhmadi-Lopez:2026dte,Verma:2025lmr} modify the same Friedmann equation with
different functions of \(T\).  Their common tests include background distances,
perturbations, local Lorentz issues, growth, and the consistency of the BAO and
SN distance scales.  Torsion-based viscous, Einstein--Cartan, and condensation
variants~\cite{Yang:2022efz,Naik:2026xpu,Mishra:2025kzu,Wu:2024weu,
Legner:2025hrt} belong to this same family when their main effect is a
torsion-driven change of the effective Friedmann dynamics.

\subsection{\texorpdfstring{\(f(Q)\)}{f(Q)} gravity}

Symmetric teleparallel gravity replaces torsion by nonmetricity.  In the
coincident formulation the nonmetricity scalar is \(Q=6H^2\) for a flat FLRW
background, and an \(f(Q)\) action takes the form
\begin{equation}
S=\frac{1}{2}\int d^4x\sqrt{-g}\,f(Q)+S_{\rm m} .
\end{equation}
The modified Friedmann equations are
\begin{equation}
\begin{aligned}
6f_QH^2-\frac{1}{2}f&=\rho,\\
\left(12H^2f_{QQ}+f_Q\right)\dot H&=-\frac12(\rho+p),
\end{aligned}
\end{equation}
where \(f_Q=df(Q)/dQ\), \(f_{QQ}=d^2f(Q)/dQ^2\), and \(\rho\) and \(p\) denote
the total energy density and pressure.  Thus the same function \(f(Q)\)
controls both the expansion rate and the response of perturbations.  A DESI-era
representative is the power-law tangent-hyperbolic model~\cite{Kavya:2025vsj},
\begin{equation}
f(Q)=Q+\epsilon Q_0\left(\frac{Q}{Q_0}\right)^\eta
\tanh\!\left(\frac{Q_0}{Q}\right),
\end{equation}
with \(Q_0\) the present value of \(Q\), and \(\epsilon\) and \(\eta\) model
parameters.  Substitution into the first Friedmann equation gives a modified
\(E(z)=H(z)/H_0\), and the model is constrained by DESI BAO, RSD, cosmic
chronometers, gravitational-wave standard sirens, SNe, and CMB distance
information.  Its relevance to the Hubble tension comes from changing the
late-time distance ratios and the growth equation at the same time, with
\(r_d\) either fitted or tied to CMB information depending on the data
combination.  Other \(f(Q)\) studies
~\cite{Lazkoz:2019sjl,Sakr:2024eee,Najera:2025htf,Solanki:2026tpi,
Quiros:2022uns,Quiros:2023owm} test the same nonmetricity mechanism.  Joint
shifts in \(H_0\) and \(S_8\) require compatibility with RSD, weak lensing, CMB
lensing, and the modified effective gravitational response.

\subsection{Other models}

The remaining modified-gravity proposals are more diverse, but their common
cosmological role is simpler.  They alter the vacuum sector, the infrared
behavior of gravity, or the number and mass of gravitational degrees of freedom,
thereby changing the effective dark-energy density or the late-time Friedmann
equation.  Curvature-squared or topological corrections, nonlocal and
vacuum-sector modifications, and infrared spin-2 or brane constructions have all
been used to test whether this freedom can absorb part of the Hubble discrepancy
~\cite{Wang:2021kuw,Petronikolou:2025mlm,Lohakare:2026jns,
DiValentino:2022eot,Bouche:2022qcv,Gangopadhyay:2022bsh,Escamilla:2024xmz,
Perez:2020cwa,LinaresCedeno:2020uxx,Landau:2022mhm,Singh:2023jxu,Das:2023rvg,
Plaza:2025nip,Dvali:2000hr,Deffayet:2001pu,Bag:2021cqm,Liu:2024dlf,
Apostolopoulos:2024muh,deRham:2010kj,Hassan:2011zd,Bassi:2023ymf,
Dwivedi:2024okk,Hogas:2025ahb,DeFelice:2022mcd,Garcia-Saenz:2025htt,
 Banik:2021woo}.  An extended uncertainty principle with a minimum
 measurable momentum, or equivalently a maximum length, has also been
 used to relate an effective photon mass to \(H_0\)~\cite{Nozari:2024wir}.  For example, bimetric gravity introduces an additional spin-2
sector whose interaction with the metric can behave as an effective dynamical
dark-energy density.  DESI-era bimetric fits can raise \(H_0\) to intermediate
values in selected BAO and SN combinations~\cite{Hogas:2025ahb}.  The shared
limitation is that the change must survive stability, gravitational-wave,
growth, and local-gravity tests.  The common conclusion is that a modified
gravitational action can connect \(H_0\), lensing, growth, and the vacuum
sector.  A restricted high-\(H_0\) fit does not establish viability without the
corresponding perturbation and screening consistency checks.  The useful
diagnostic is therefore not the fitted \(H_0\) alone but whether the same
modified field equations give a consistent background, perturbation, lensing,
growth, and screening limit.

\section{Early-time solutions}\label{sec:early}

Early-time explanations act on the standard-ruler side of the Hubble tension.  The
CMB does not measure \(H_0\) directly.  It fixes the angular acoustic scale
\(\theta_s=r_s(z_\ast)/D_A(z_\ast)\), together with the physical matter and
baryon densities~\cite{Hu:2001bc,Planck:2018vyg}.  BAO then carries the same
standard ruler to low redshift through the drag-epoch ruler \(r_d\)
~\cite{Bernal:2016gxb,Aylor:2018drw,Knox:2019rjx}.  A model that increases the
expansion rate before recombination can therefore reduce the standard ruler
before CMB and BAO distances are inferred.  In this framework the two
relevant rulers are
\begin{equation}
r_s(z_\ast)=\int_{z_\ast}^{\infty}\frac{c_s(z)}{H(z)}\,dz,\qquad
r_d=\int_{z_d}^{\infty}\frac{c_s(z)}{H(z)}\,dz .
\end{equation}
Here \(c_s(z)\) is the photon-baryon sound speed, \(z_\ast\) is the
last-scattering redshift, and \(z_d\) is the baryon-drag redshift.
A temporary contribution to \(H(z)\) before recombination decreases these
integrals.  The same contribution must then fade rapidly, because the CMB peak
structure, the diffusion damping scale, CMB lensing, BBN, BAO consistency,
SN distances, and the growth of structure all constrain any residual
change in the background or perturbations.  Reducing \(r_s\) or \(r_d\) is
therefore a necessary direction for many early-time models but not a sufficient
condition for a consistent fit~\cite{Schoneberg:2021qvd,Huang:2024erq,
Krishnan:2020obg,Jedamzik:2020zmd,Vagnozzi:2021gjh,Pedrotti:2024kpn}.

\begin{figure}
\centering
\includegraphics[width=0.92\linewidth]{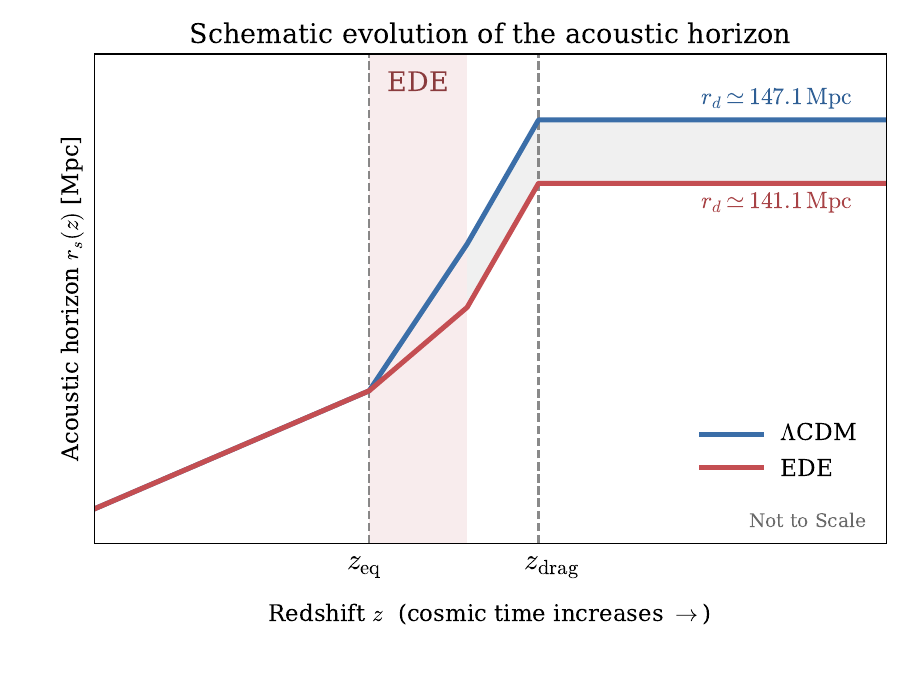}
\caption{
Schematic illustration of the response of the sound horizons \(r_s\) and \(r_d\)
to a transient pre-recombination increase in \(H(z)\).  The numerical values are
representative values from Planck, DESI, ACT, and EDE analyses assembled for
schematic comparison~\cite{Planck:2018vyg,DESI:2025zgx,Poulin:2025nfb}.
}
\label{fig:acoustic_horizon_schematic}
\end{figure}

Fig.~\ref{fig:acoustic_horizon_schematic} illustrates the standard-ruler
mechanism behind early-time solutions.  A short-lived component near
matter-radiation equality raises \(H(z)\) during the part of the integral that
contributes significantly to \(r_s\) and \(r_d\).  For a fixed observed
\(\theta_s\), the smaller ruler is matched by a smaller \(D_A(z_\ast)\), which
corresponds to a larger inferred \(H_0\).  This mechanism also explains why the
timing of the extra component is observationally constrained.  If the injection happens too early
or too late, it becomes less efficient at reducing the acoustic ruler and more
visible in other precision observables.

\subsection{Early dark energy}

The axion-like EDE model gives a concrete realization of this short-lived energy
injection.  In the standard implementation~\cite{Poulin:2018cxd,Smith:2019ihp},
a scalar field \(\phi\) is initially frozen by Hubble friction on a potential of
the form
\begin{equation}
V(\phi)=V_0\left[1-\cos\left(\frac{\phi}{f}\right)\right]^n ,
\end{equation}
where \(V_0\) sets the potential scale and \(f\) is the axion decay constant.
The field behaves approximately as vacuum energy while it is frozen.  When
\(H\) falls below the effective mass scale of the potential, at a critical
redshift \(z_c=a_c^{-1}-1\), the field rolls and then oscillates.  The parameter
\(\theta_i=\phi_i/f\) fixes the initial displacement, \(f_{\rm EDE}\) denotes
the maximum fractional contribution of the field to the total energy density,
and \(n\) controls how the field redshifts after the transition.  A useful
effective-fluid description gives
\begin{align}
\Omega_\phi(a)&=
\frac{2\Omega_\phi(a_c)}
{(a/a_c)^{3(1+w_n)}+1},\\
w_\phi(a)&=\frac{1+w_n}{1+(a_c/a)^{3(1+w_n)}}-1 ,
\end{align}
with the cycle-averaged equation of state
\begin{equation}
\begin{split}
w_n&=\frac{n-1}{n+1},\\
\rho_\phi&\propto a^{-3(1+w_n)}=a^{-6n/(n+1)}
\quad (a\gg a_c).
\end{split}
\end{equation}
Thus \(n=1\) gives matter-like dilution, \(n=2\) gives radiation-like dilution,
\(n=3\) dilutes faster than radiation, and \(n\rightarrow\infty\) approaches a
kinetic-energy-dominated \(a^{-6}\) scaling.  Rapid dilution is essential
because the field must lower the sound horizon before recombination while
leaving only a small residual contribution afterwards.  Otherwise the same
energy component would remain visible in the CMB peak pattern, diffusion
damping, lensing reconstruction, BAO distances, and late-time growth.  EDE is
therefore a benchmark pre-recombination mechanism only when the high-\(H_0\)
direction is tested against the full CMB+BAO+SNe+growth consistency network.

Modern BAO and SN data make this mechanism directly testable.  DESI DR2
BAO measurements fix the low-redshift standard-ruler combination with high
precision~\cite{DESI:2025zgx}, while PantheonPlus and DES-Dovekie SN samples test
the associated distance-redshift relation~\cite{Scolnic:2021amr,Brout:2022vxf,
DES:2025sig}.  The question is therefore whether a smaller \(r_d\), generated by
EDE before recombination, can be embedded in a coherent CMB+BAO+SNe fit
~\cite{Qu:2024lpx,Jiang:2025hco}.  The reference constraints in this review use
the code and likelihood setup summarized in Appendix~\ref{app:ede_methodology}.
In the main-text figures and the appendix table, Planck denotes the Planck-only
likelihood block, CMB denotes the same block with ACT DR6 CMB lensing, PP
denotes PantheonPlus without local calibration, \(+\mathrm{SH0ES}\) denotes the
PantheonPlus likelihood with SH0ES calibration, DESI denotes DESI DR2 BAO, and
DES denotes the DES-Dovekie SN likelihood.  The \(+H_0\) label is reserved for
the direct local \(H_0\) measurement added to the DES-Dovekie combination.
These calculations are intended
as benchmark constraints for comparing parameter directions across recent CMB,
BAO, and SN combinations.

\begin{figure}[htbp]
\centering
\includegraphics[width=\linewidth,keepaspectratio]{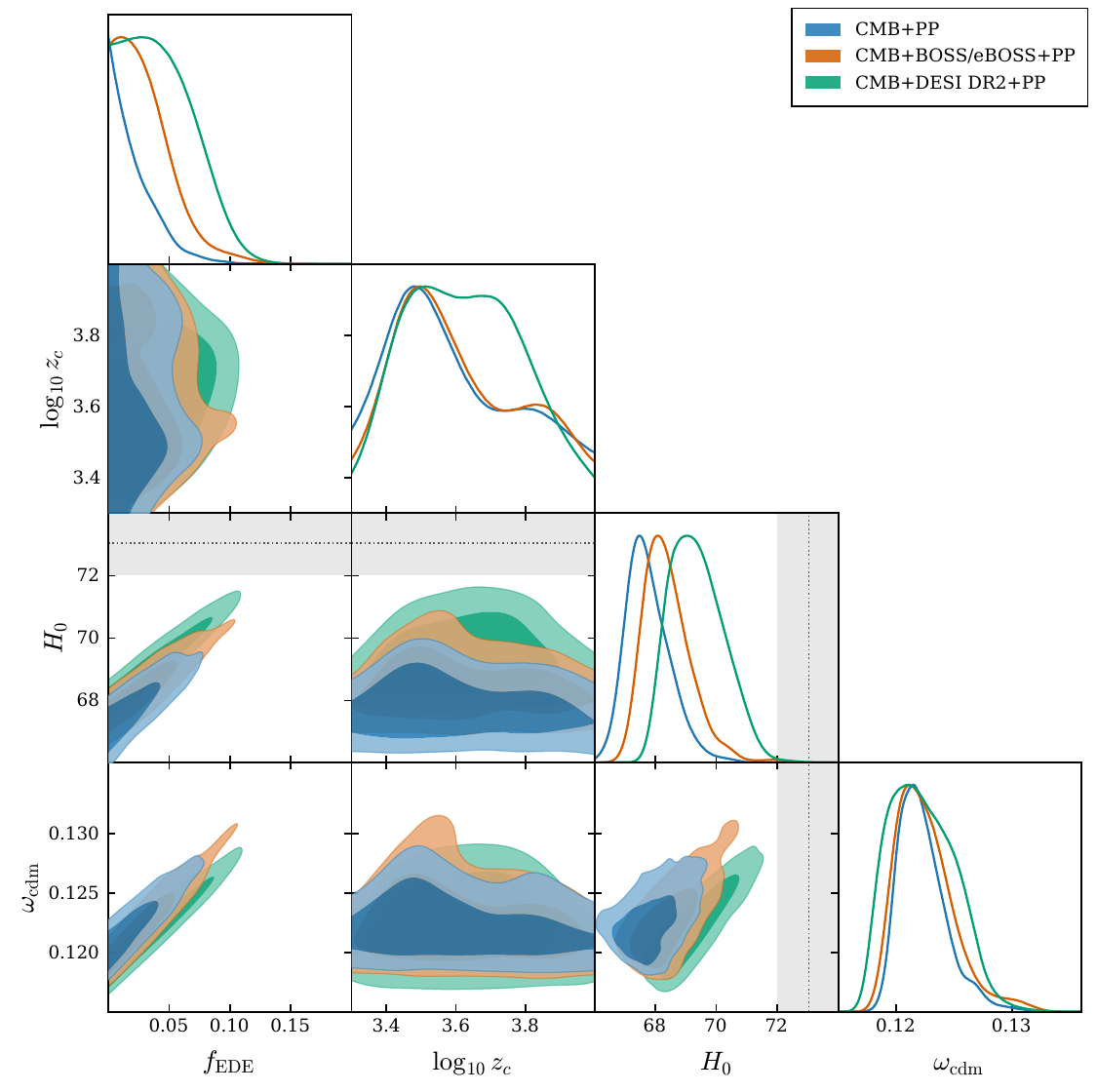}
\caption{
Reference fixed-$n$ EDE constraints from representative DESI-era CMB, BAO, and
PantheonPlus combinations.  The displayed parameters are the EDE peak fraction
$f_{\rm EDE}$, the critical redshift $\log_{10}z_c$, the Hubble constant $H_0$,
and the physical cold-dark-matter density $\omega_{\rm cdm}$.  The gray band
marks the local distance-ladder reference range used in this set of EDE figures.
}
\label{fig:ede_fixedn_benchmark}
\end{figure}

The fixed-$n$ benchmark constraints in Fig.~\ref{fig:ede_fixedn_benchmark}
show how the standard-ruler mechanism appears in parameter space.  Without
local calibration, the CMB+PP and CMB+BOSS/eBOSS+PP posteriors remain
concentrated near the low-\(f_{\rm EDE}\) boundary and lower \(H_0\), while
CMB+DESI DR2+PP allows a somewhat broader extension toward higher values.  The
marginalized means across these uncalibrated PP combinations span
\(f_{\rm EDE}\simeq0.02\)--\(0.04\) and
\(H_0\simeq67.8\)--\(69.4\,{\rm km\,s^{-1}\,Mpc^{-1}}\).  Within each posterior,
the higher-\(f_{\rm EDE}\) direction remains correlated with higher \(H_0\) and
\(\omega_{\rm cdm}\), as expected for EDE fits.  The smaller sound horizon raises the
inferred \(H_0\), while the CMB fit compensates through the matter density and
the detailed acoustic spectrum.  The differences among the BOSS/eBOSS and
DESI DR2 contours show how BAO data constrain the allowed ruler shift, but these
uncalibrated combinations alone do not select the SH0ES-like high-\(H_0\)
region.  This pattern is
consistent with recent literature.  Profile-likelihood analyses
can find high-likelihood regions near \(f_{\rm EDE}\simeq0.09\) and
\(H_0\simeq71\,{\rm km\,s^{-1}\,Mpc^{-1}}\) without a SH0ES prior
~\cite{Poulin:2025nfb}, whereas ACT, SPT, and combined CMB+DESI analyses show a
more data- and statistic-dependent picture
~\cite{AtacamaCosmologyTelescope:2025nti,SPT-3G:2025vyw,Toda:2025kcq,Wang:2025djw}.

\begin{figure}[htbp]
\centering
\includegraphics[width=\linewidth,keepaspectratio]{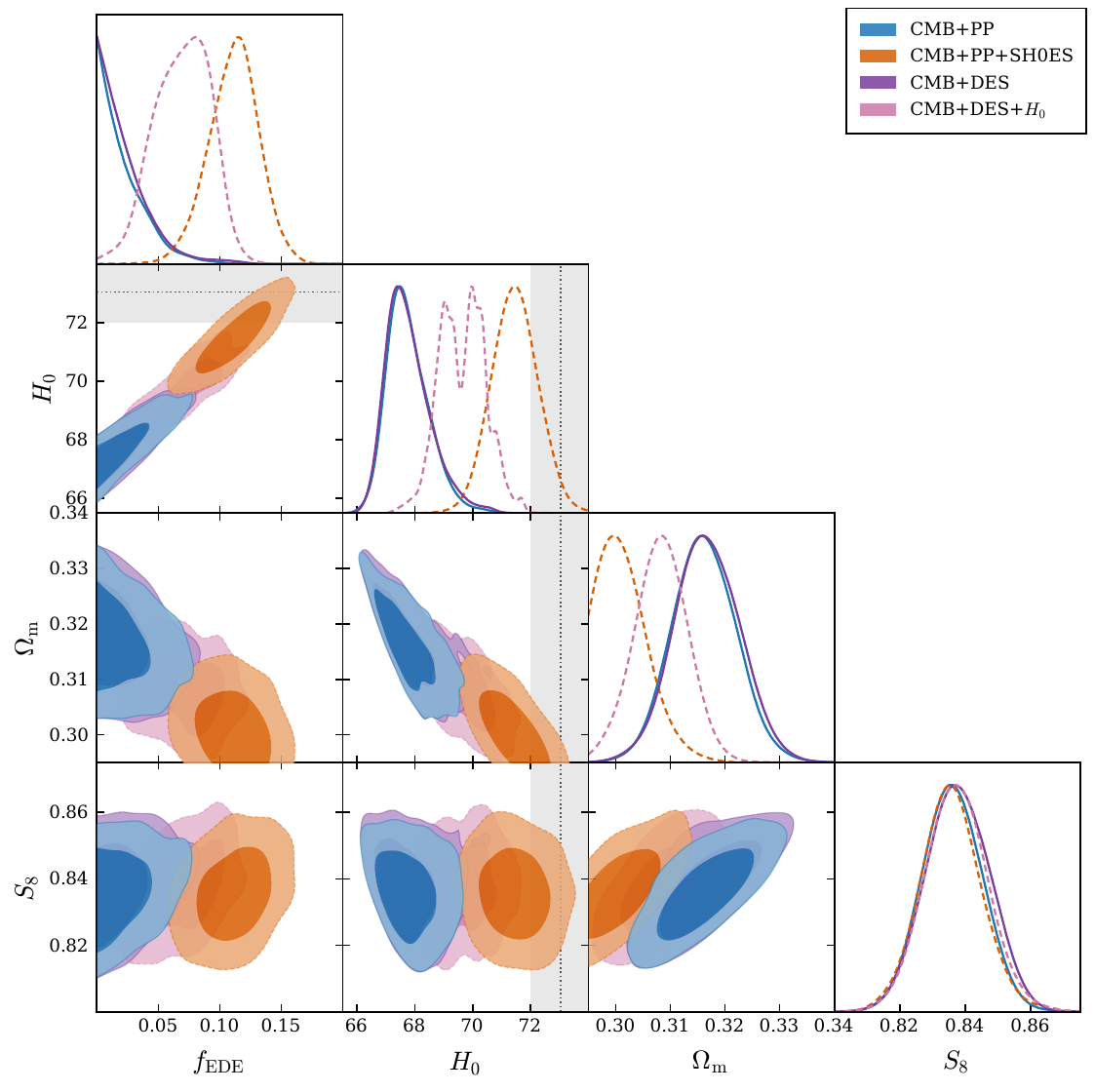}
\caption{
DES-Dovekie comparison for fixed-$n$ EDE.  Here DES denotes the DES-Dovekie
likelihood.  The plotted parameters are \(f_{\rm EDE}\), \(H_0\), \(\Omega_{\rm m}\),
and \(S_8\).  Contours are shown for CMB+PP, CMB+PP+SH0ES, CMB+DES, and
CMB+DES+\(H_0\).
}
\label{fig:ede_dovekie}
\end{figure}

The SN data choice tests another part of the same consistency
condition.  Fig.~\ref{fig:ede_dovekie} compares CMB+PP and
CMB+PP+SH0ES with CMB+DES and CMB+DES+\(H_0\).  The uncalibrated CMB+PP and
CMB+DES posteriors both
favor low \(f_{\rm EDE}\) and \(H_0\simeq68\,{\rm km\,s^{-1}\,Mpc^{-1}}\).
Adding the direct local \(H_0\) measurement to the DES combination shifts the
posterior to the intermediate region \(f_{\rm EDE}\simeq0.07\) and
\(H_0\simeq69.6\,{\rm km\,s^{-1}\,Mpc^{-1}}\), which remains distinct from the
SH0ES-calibrated PantheonPlus result.  The comparison shows that a smaller
pre-recombination ruler has to remain compatible with the SN Hubble
diagram.  Sound-horizon-independent probes based on SNe, galaxy surveys,
and standard sirens provide complementary tests through observables that are
partly independent of the same \(r_d\) standard ruler
~\cite{Philcox:2022sgj,Kable:2024mgl,Giare:2024syw}.

The BAO and SN comparisons do not exhaust the test of EDE.  The same
energy injection that reduces \(r_s\) also changes the projected equality scale,
the ratio of the acoustic scale to the diffusion damping scale, the peak
heights, CMB lensing, and the matter power spectrum.  Raising \(f_{\rm EDE}\)
therefore requires correlated shifts in \(\omega_{\rm cdm}\), \(n_s\), lensing,
and growth observables~\cite{Hill:2020osr,Ivanov:2020ril,DAmico:2020ods,Pedrotti:2024kpn}.
Large-scale-structure (LSS) and weak-lensing information therefore put strong pressure
on the earliest high-\(H_0\) EDE regions
~\cite{Hill:2020osr,Ivanov:2020ril,DAmico:2020ods}, while alternative covariance
choices, profile-likelihood treatments, and EDE variants can leave viable
high-likelihood islands~\cite{Smith:2020rxx,Niedermann:2020qbw,Herold:2022iib,
Cruz:2023cxy}.

\begin{figure}[htbp]
\centering
\includegraphics[width=\linewidth,keepaspectratio]{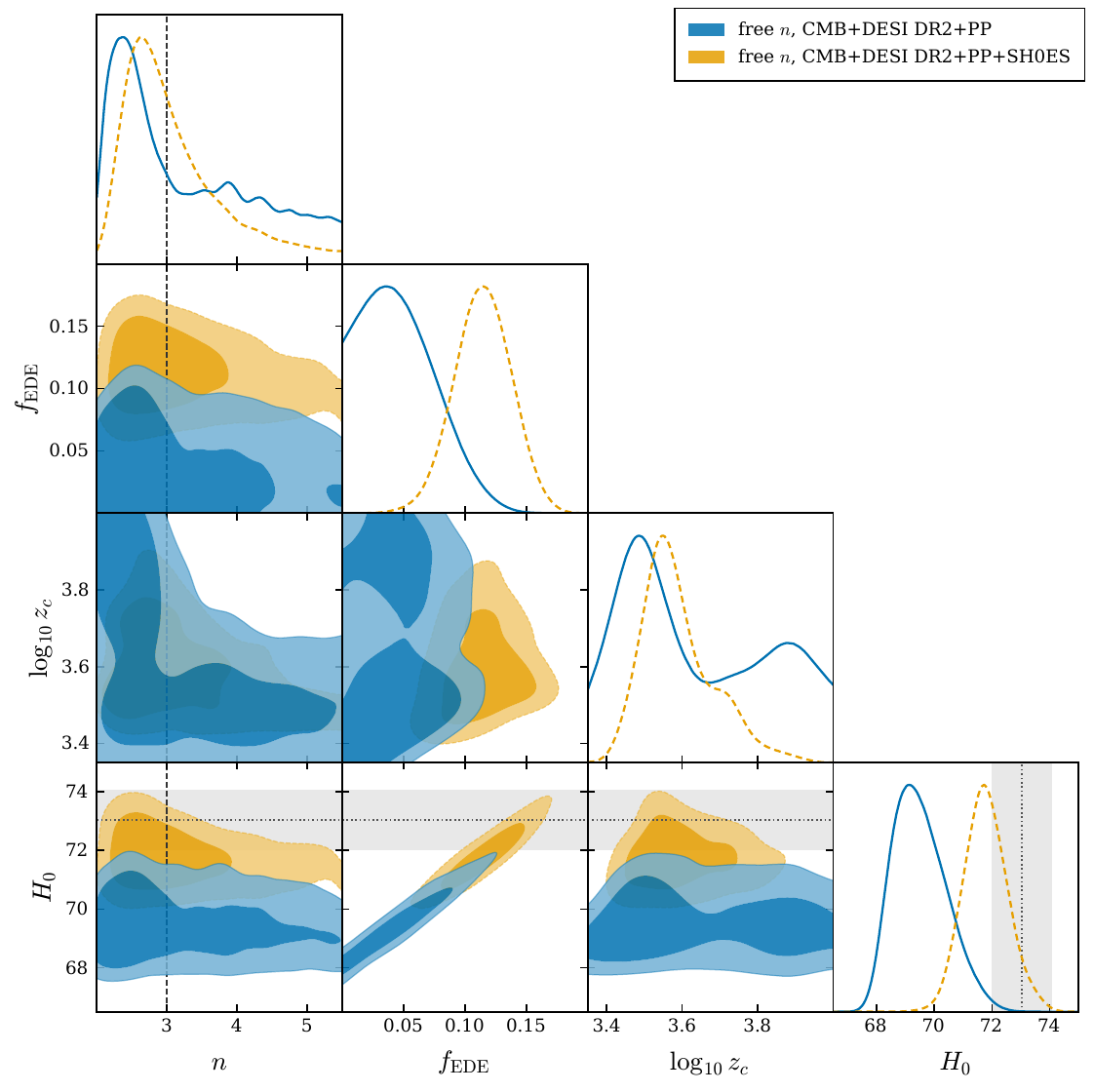}
\caption{
Free-$n$ EDE constraints for CMB+DESI DR2+PP, with and without SH0ES
calibration.  The dashed vertical line marks the fixed $n=3$
reference model used in the fixed-potential benchmark calculation.  The plotted
parameters are \(n\), \(f_{\rm EDE}\), \(\log_{10}z_c\), and \(H_0\).
}
\label{fig:ede_free_n}
\end{figure}

The potential exponent \(n\) is a useful way to display this CMB response.
Larger \(n\) makes the post-oscillation density dilute more rapidly, but it also
changes the time profile of the energy injection and the perturbation response
around recombination.  The original axion-like analysis found \(n=3\) to be a
good compromise because it can reduce \(r_s\) efficiently while keeping the
ratio between the sound horizon and the diffusion damping scale, the peak
heights, and the equality imprint closer to the observed CMB spectra than the
\(n=2\) or very large-\(n\) limits
~\cite{Poulin:2018cxd}.  The free-\(n\) reference chains in
Fig.~\ref{fig:ede_free_n} follow the same logic.  They do not show a preference
for arbitrarily rapid post-transition dilution in the combined CMB+BAO+SNe
information.  Without SH0ES calibration the free-\(n\) posterior remains at
lower \(H_0\) and \(f_{\rm EDE}\), whereas adding SH0ES shifts it toward
\(H_0\simeq71.8\,{\rm km\,s^{-1}\,Mpc^{-1}}\) and
\(f_{\rm EDE}\simeq0.11\).  The latter high-\(H_0\) direction remains tied to a
finite EDE fraction and to the same correlated shifts in \(z_c\),
\(\omega_{\rm cdm}\), and the CMB spectrum that appear in the fixed-\(n\)
model.  High-resolution CMB data
therefore sharpen EDE constraints through polarization, damping, lensing
reconstruction, and peak residuals~\cite{Poulin:2021bjr,Smith:2022hwi,
Jiang:2022uyg,LaPosta:2021pgm,Smith:2023oop,Ye:2023zel}.

A related strategy is to make the field dissipate more efficiently after it has
reduced the acoustic ruler.
Anti-de Sitter early dark energy (AdS-EDE) provides a representative example~\cite{Ye:2020btb}.  In the
\(\phi^4+\)AdS construction, the scalar potential is modeled as
\begin{equation}
V(\phi)=
\begin{cases}
V_0\phi^4-V_{\rm ads},
& \phi<\left(\dfrac{V_{\rm ads}}{V_0}\right)^{1/4},\\[6pt]
0,
& \phi>\left(\dfrac{V_{\rm ads}}{V_0}\right)^{1/4},
\end{cases}
\end{equation}
where \(V_{\rm ads}\) denotes the depth of the negative-potential region.  A
small positive late-time vacuum energy is added separately in the
cosmological background.  The field starts frozen on the side of the potential,
rolls when \(H^2\simeq \partial_\phi^2V\), and then crosses a region where
\(V<0\).  During this AdS phase,
\(\rho_\phi=\dot{\phi}^2/2+V\) and
\(P_\phi=\dot{\phi}^2/2-V\) give \(w_\phi=P_\phi/\rho_\phi>1\), so the scalar
energy density redshifts faster than in an ordinary oscillating EDE model.
This rapid dissipation enables a sizable pre-recombination energy
injection while leaving a smaller residual fraction at recombination, where CMB
peak-structure and damping bounds are particularly tight.  The model remains a
specific dynamical proposal with its own background-stability and parameter
choices, but it illustrates the broader direction pursued by AdS-EDE
extensions, chain or multi-field EDE, acoustic dark energy, and new early dark energy (NEDE)-like
triggers~\cite{Ye:2020oix,Jiang:2021bab,Freese:2021rjq,Bella:2026zuk,
Lin:2019qug,Lin:2020jcb,Yin:2020dwl,Niedermann:2019olb,Niedermann:2020dwg}.

These variants address part of the phenomenological pressure, while leaving the
basic model-building cost in place.  Standard axion-like EDE must arrange a
mass scale and initial displacement so that the field becomes important close
to equality or recombination, precisely where a reduction of the acoustic ruler
is useful~\cite{Poulin:2018cxd,Smith:2019ihp,Hill:2020osr,Vagnozzi:2023nrq}.
This is the EDE coincidence and tuning problem.  The same issue reappears in
triggered models in a different form.  Massive-neutrino or neutrino-assisted
EDE~\cite{Sakstein:2019fmf,CarrilloGonzalez:2023lma}, dark-matter triggers
~\cite{Lin:2022phm}, coupled EDE~\cite{Trodden:2022zye}, and phase-transition
NEDE~\cite{Niedermann:2021vgd,Niedermann:2021ijp,Cruz:2022oqk,Cruz:2023cxy,
Cruz:2023lmn,Chatrchyan:2024xjj,Garny:2025kqj} try to replace a tuned scalar
mass by a dynamical event.  The trigger sector, couplings, stability, and
perturbation evolution then become part of the observational test.

\subsection{Other models}

\begin{figure*}
\centering
\includegraphics[width=0.92\textwidth,height=0.68\textheight,keepaspectratio]{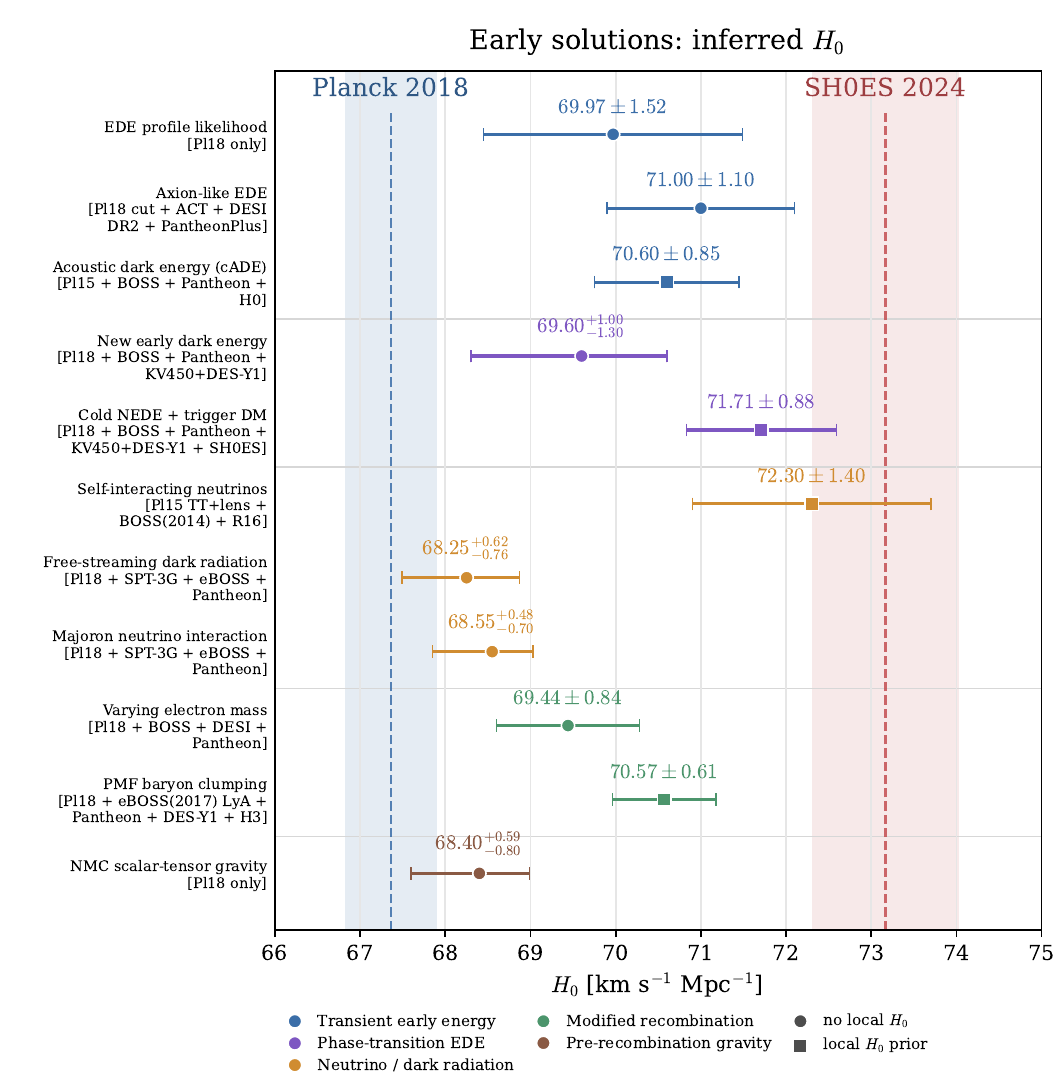}
\caption{
Representative \(H_0\) values in early-time proposals.  Circles denote analyses
without a direct local distance-ladder prior in the quoted result, while squares
denote analyses that include such a prior.  The values are compiled from
Refs.~\cite{Herold:2022iib,Poulin:2025nfb,Lin:2019qug,Niedermann:2020dwg,
Cruz:2023lmn,Kreisch:2019yzn,Khalife:2023qbu,Seto:2024cgo,Jedamzik:2020krr,
Braglia:2020iik}; the underlying data combinations and statistical procedures
are heterogeneous.
}
\label{fig:early_solutions_h0}
\end{figure*}

Fig.~\ref{fig:early_solutions_h0} places EDE in a broader landscape of
early-time proposals.  The entries constitute a heterogeneous set because the
analyses differ in data choices, priors, and whether local distance-ladder
information is included.  They nevertheless show the common target.  Models that
act before recombination seek to reduce the standard ruler while preserving
the rest of the CMB and late-time distance information.  Axion-like EDE,
especially in profile-likelihood treatments, can reach
\(H_0\simeq70\)--\(71\,{\rm km\,s^{-1}\,Mpc^{-1}}\) without using a direct local
distance-ladder prior in the quoted Planck-only or CMB/DESI combinations
~\cite{Poulin:2018cxd,Herold:2022iib,Poulin:2025nfb}.  Here ``Pl cut'' denotes
the restricted Planck Release 3 (PR3) likelihood used in the Planck-ACT (P-ACT) construction of
Ref.~\cite{Poulin:2025nfb}, with TT limited to \(\ell<1000\) and TE/EE limited
to \(\ell<600\) before combination with ACT.

Other early-time proposals modify different ingredients of the
pre-recombination calculation.  Extra dark radiation or a larger \(N_{\rm eff}\)
increases the expansion rate and can reduce the sound horizon, but free-streaming
radiation also affects the CMB phase shift, BBN light-element abundances, and
the damping tail.  Self-interacting neutrinos, sterile-neutrino interactions,
and neutrino--dark-radiation sectors therefore try to modify both the background
 expansion and perturbation propagation~\cite{Kreisch:2019yzn,Allali:2021azp,
 Khalife:2023qbu,Buen-Abad:2024tlb,RoyChoudhury:2020dmd}.  Recombination-sector models change the
visibility function directly.  Varying
electron-mass scenarios alter atomic binding energies and the Thomson-scattering
history~\cite{Seto:2024cgo,Toda:2025kcq,Chluba:2023xqj,Toda:2025dzd,Schoneberg:2024ynd}, while primordial-magnetic-field-induced
baryon clumping changes recombination through spatial inhomogeneity
~\cite{Jedamzik:2020krr,Jedamzik:2023csc,Thiele:2021okz,Galli:2021mxk}.  Because these mechanisms also alter the damping tail,
BBN consistency, and high-resolution CMB spectra, their viability is controlled
by more than the posterior shift in \(H_0\).  Early modifications of the effective Planck mass,
such as the scalar-tensor entry shown in Fig.~\ref{fig:early_solutions_h0},
change gravitational dynamics near equality through the effective strength of
gravity~\cite{Braglia:2020iik,FrancoAbellan:2023gec}; related pure-gravity
modifications are discussed again in Sec.~\ref{sec:modified_gravity}.  In the
primordial-magnetic-field entry, \(H3\) follows the notation of
Ref.~\cite{Jedamzik:2020krr} and denotes the combined use of SH0ES, the
Megamaser Cosmology Project, and H0LiCOW.

The synthesis of this section is that early-time proposals directly address
the acoustic ruler, while the same data strongly constrain the allowed
implementation.  EDE remains the most developed realization of this strategy,
but acoustic peaks, damping scale, lensing amplitude, BAO ruler, SN
distances, BBN, and growth of structure must all be fitted simultaneously.
Post-DESI and post-ACT analyses have identified viable EDE regions in some
likelihood treatments, while also showing that an early-time explanation must be
evaluated as a full cosmological model, not merely as a one-parameter reduction
of the sound horizon~\cite{Poulin:2023lkg,Giare:2026tyk,Jedamzik:2020zmd,Pedrotti:2024kpn}.

\section{Late-time solutions}\label{sec:late}

Late-time explanations begin after the acoustic ruler has been set by
pre-recombination physics.  They therefore act on the distance-redshift
relation, on the recent composition of the dark sector, or on the local matter
distribution sampled by nearby distance indicators.  The difficulty is that these
quantities are already tied together by several measurements.  The CMB fixes the
angular acoustic scale, BAO propagates the drag-epoch ruler \(r_d\) to low
redshift, and SNe determine the relative Hubble diagram.  In
inverse-distance-ladder analyses
~\cite{Bernal:2016gxb,Aylor:2018drw,Knox:2019rjx,Schoneberg:2022ggi,Poulin:2024ken}, and in broader consistency
tests of proposed solutions~\cite{Schoneberg:2021qvd,Huang:2024erq}, these
observables leave little room for an arbitrary change of the late expansion
history if the standard ruler remains fixed.  A viable late-time model must
therefore specify how \(D_L(z)\), \(D_A(z)\), BAO distances, the CMB distance to
last scattering, and growth-related observables move together.

\begin{figure}
\centering
\includegraphics[width=0.86\linewidth]{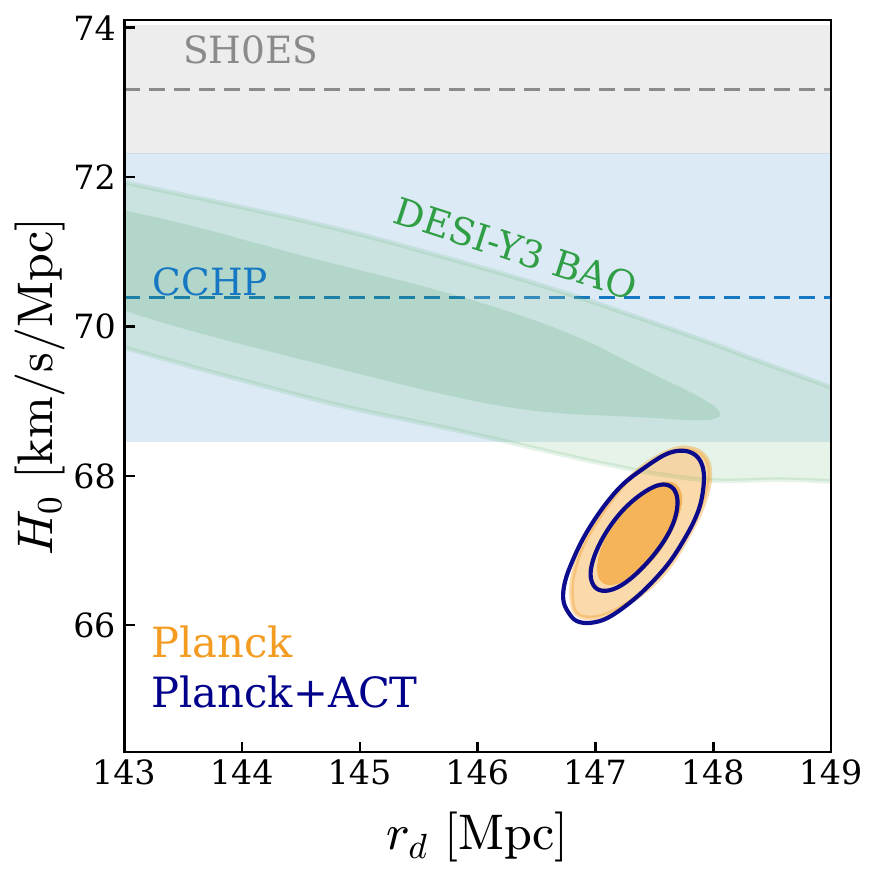}
\caption{
The \(H_0\)--\(r_d\) plane for the Hubble tension.  The plot summarizes the
standard-ruler consistency condition discussed in
Refs.~\cite{Bernal:2016gxb,Aylor:2018drw,Knox:2019rjx,Keeley:2022ojz,Kanodia:2025jqh}.
}
\label{fig:h0rd_late}
\end{figure}

Fig.~\ref{fig:h0rd_late} shows the same restriction in the
\(H_0\)--\(r_d\) plane.  Late-time physics can change
distances at \(z\lesssim {\cal O}(1)\), but BAO and SNe do not allow the
horizontal and vertical directions in this plane to vary independently.  Many
low-redshift explanations are therefore framed as consistency or no-go tests
~\cite{Keeley:2022ojz,Huang:2024erq,Bansal:2026axl}.  A larger fitted \(H_0\)
inside one parameterization is informative only if the same parameterization
keeps the CMB, BAO, and SN likelihoods consistent.  The DESI DR2 and
megamaser distance-duality analysis of Ref.~\cite{Kanodia:2025jqh} makes this
point in a calibration-independent form, finding that the early and late
distance scales remain hard to reconcile when \(r_d\) is held fixed.  This
agrees with the broader conclusion that early-time and late-time deformations
face different, but comparably sharp, consistency tests~\cite{Vagnozzi:2023nrq}.

\subsection{Dynamical dark energy}

The minimal homogeneous starting point is a dark-energy equation of state
different from a cosmological constant.  With
\(E(z)\equiv H(z)/H_0\), a standard late-time parametrization can be written as
\begin{equation}
\begin{aligned}
E^2(z)=&\,\Omega_{\rm m}(1+z)^3+\Omega_{\rm r}(1+z)^4
\\
&+\Omega_k(1+z)^2+\Omega_{\rm DE}X(z),\\
X(z)=&\,\exp\!\left[3\int_0^z\frac{1+w(z')}{1+z'}\,dz'\right],
\end{aligned}
\end{equation}
where \(\Omega_{\rm m}\), \(\Omega_{\rm r}\), \(\Omega_k\), and
\(\Omega_{\rm DE}\) denote the present-day matter, radiation, curvature, and
dark-energy density parameters, while \(X(z)\) encodes the dark-energy density
evolution.  A change in \(w(z)\) changes \(X(z)\), and therefore alters
\(H(z)\), \(D_L(z)\), \(D_A(z)\), and the CMB distance to last scattering.
Constant-\(w\) and
CPL parameterizations
~\cite{Chevallier:2000qy,Linder:2002et} show this effect with only one or two
 additional parameters.  Scalar-field and two-field
 realizations~\cite{Copeland:2006wr,Panpanich:2019fxq,Fu:2023tlp,Roy:2023vxk}
 give possible physical origins to non-\(\Lambda\)CDM expansion histories.
 Related late-time proposals invoke slow-rolling scalar dynamics,
 viscous dark-energy production, or gravitational particle production to alter
 the expansion history or its inferred normalization
 ~\cite{Montani:2023ywn,Navone:2025gxr,Erdem:2024vsr,Erdem:2025xtr}.
 Addressing the Hubble tension requires more than demonstrating
that a non-\(\Lambda\) model provides a better fit.  If \(r_d\) is unchanged, BAO and
SNe restrict the allowed \(w(z)\) history to a narrow allowed range
~\cite{Vagnozzi:2019ezj,Alestas:2021xes}.  DESI BAO measurements
~\cite{DESI:2024mwx,DESI:2025zgx} sharpen this test in combination with CMB and
SN data, producing preferences for evolving dark-energy
parameterizations in some analyses.  Their relevance for the Hubble tension
depends on whether the same time dependence raises \(H_0\) while preserving the
BAO ruler, the SN Hubble diagram, growth information, and the CMB
distance.  The inferred evidence for evolving dark energy can itself change
when local-\(H_0\) information is included or excluded from the analysis
~\cite{Pang:2025lvh}, and apparent improvements remain sensitive to data choice
 and functional form~\cite{Zhang:2024eba,Efstathiou:2024xcq,Huang:2025som,
 DES:2025tir,Popovic:2025glk,RoyChoudhury:2024wri,RoyChoudhury:2025dhe,
 RoyChoudhury:2025iis}.  Redshift-binned SN analyses have also examined
 apparent evolution in the fitted \(H_0\), with the interpretation remaining
 sensitive to sample construction and model assumptions
 ~\cite{Dainotti:2025qxz,Dainotti:2021pqg,Dainotti:2022bzg,DeSimone:2024lvy}.

A representative late-transition model with an explicit physical origin is
Parker's vacuum metamorphosis
~\cite{DiValentino:2017rcr,DiValentino:2020kha}.  In this model the transition
is tied to nonperturbative vacuum effects of a massive scalar field in curved
spacetime and has a specified curvature trigger.
When the Ricci scalar reaches the scale set by the field mass,
\begin{equation}
R=6(\dot H+2H^2+k/a^2)=m^2 ,
\end{equation}
where \(a\) is the scale factor, \(k\) is the spatial-curvature parameter, and
\(m\) is the scalar-field mass.
At this point the vacuum state changes and the curvature is driven toward
\(R=m^2\).  With
\(M=m^2/(12H_0^2)\), the post-transition expansion in the simplest form is
approximately
\begin{equation}
E^2(z)=(1-M-\Omega_k)(1+z)^4+\Omega_k(1+z)^2+M ,
\qquad z\le z_t .
\end{equation}
Here \(z_t\) is the transition redshift.
The effective late component has \(w_{\rm eff}<-1\) after the transition and can
raise the CMB-inferred \(H_0\) without adding more free parameters than
\(\Lambda\)CDM in the original version.  The same mechanism also changes
\(\Omega_{\rm m}\), distances, and structure growth.  Vacuum-metamorphosis fits that
give high \(H_0\) from CMB or CMB+BAO combinations are disfavored in joint fits
once multiple distance probes, especially SNe, are included
~\cite{DiValentino:2020kha}.

Phenomenological late dark energy transitions clarify why this happens.  In the
step model studied by Benevento, Hu, and Raveri~\cite{Benevento:2020fev}, the
dark-energy density is modified as
\(\rho_{\rm DE}(z)=[1+f(z)]\tilde{\rho}_\Lambda\), with a smooth step described
by an amplitude, a transition redshift, and a width.  If the SH0ES measurement
is used only as a prior on \(H_0\) at \(z=0\), a very low-redshift transition
can raise the fitted value to the local range while leaving CMB and BAO nearly
unchanged.  The interpretation changes when the local distance ladder is treated
as a calibration of the absolute magnitude of SNe extending into the
Hubble flow.  In that test the allowed \(H_0\) shift is much smaller, because
the transition must also fit the calibrated SN Hubble diagram.  For
related late-transition and vacuum-sector parameterizations
~\cite{Alestas:2021luu,Alestas:2020zol,Li:2019yem,Hernandez-Almada:2020uyr,
Yang:2021eud,Benaoum:2020qsi,Banihashemi:2020wtb,SolaPeracaula:2023swx,
Yarahmadi:2024oqv,Normann:2021bjy}, the same criterion is central.  These
models probe how much late-time freedom remains, while the standard-ruler and
SN calibration constraints remain the limiting observables.

\subsection{Interacting dark energy}

Dark-sector energy transfer changes the late-time densities more directly.  It
does not modify the visible-sector gravitational action, but it changes the
continuity equations that determine \(\rho_{\rm c}\), \(\rho_{\rm DE}\), and
\(H(z)\).  A representative interacting-dark-energy parameterization is
\begin{equation}
\dot{\rho}_{\rm c}+3H\rho_{\rm c}=Q,\qquad
\dot{\rho}_{\rm DE}+3H(1+w)\rho_{\rm DE}=-Q ,
\end{equation}
where the sign and time dependence of \(Q\) determine the direction of energy
flow.  An interaction such as \(Q={\cal H}\xi\rho_{\rm DE}\), with
\({\cal H}\) the conformal Hubble rate and \(\xi\) a coupling constant, can lower the inferred
matter density and raise the fitted \(H_0\) in Planck+DESI combinations
~\cite{Giare:2024smz}.  The same interaction also changes the perturbation
equations.  A viable interacting model must specify its perturbative stability,
rest-frame sound speed, momentum-transfer prescription, and treatment of
early-time large-scale instabilities~\cite{Valiviita:2008iv,Li:2014eha,Dai:2019vif}.
Growth, lensing, and clustering data therefore become as important
as the background distances.  In the DESI-era analysis, the region that weakens
the SH0ES discrepancy can imply a low \(\Omega_{\rm m}\) and a high \(\sigma_8\),
which shifts the pressure to large-scale-structure tests.  Earlier coupled
vacuum fits show the same pattern~\cite{DiValentino:2019jae}.  Energy exchange
can accommodate a higher local distance-ladder value, but its success depends on the
full CMB, BAO, SNe, growth, and direct-calibrator combination.  Recent DESI-era
IDE constraints should therefore be read as joint background-and-perturbation
tests rather than as evidence for a purely geometric correction
~\cite{Silva:2025hxw,Benisty:2024lmj,Pan:2025qwy}.

\begin{figure*}[t]
\centering
\includegraphics[width=0.92\textwidth,height=0.68\textheight,keepaspectratio]{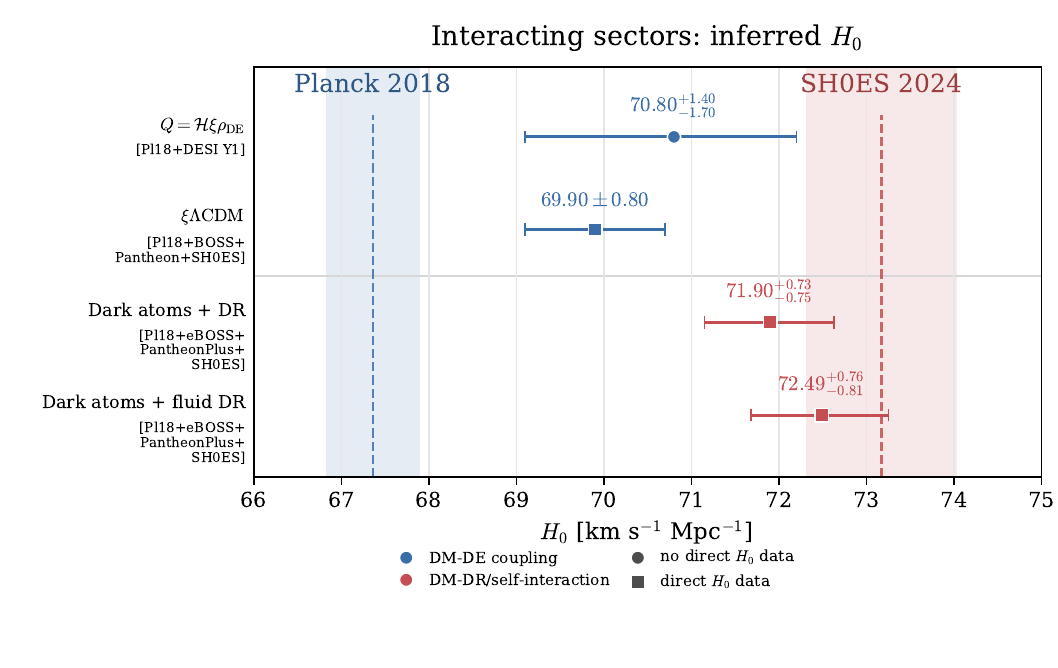}
\caption{
Representative \(H_0\) values in interacting dark-sector proposals.  Circles
denote fits without a direct local distance-ladder or analogous \(H_0\) input,
while squares denote data combinations that include a direct distance-ladder or
absolute-magnitude calibrator.  The plotted entries include direct energy
exchange, DM--DE coupling, and DM--DR or self-interaction scenarios, compiled
from Refs.~\cite{Giare:2024smz,DiValentino:2019jae,Buen-Abad:2024tlb}.
}
\label{fig:interacting_h0}
\end{figure*}

Fig.~\ref{fig:interacting_h0} summarizes representative \(H_0\) values for
direct interacting dark-sector proposals.

\subsection{Local inhomogeneity}

The last late-time route changes the interpretation of the local expansion
field.  Nearby distance indicators may sample a local region whose effective
expansion differs from the asymptotic Friedmann--Lema\^itre--Robertson--Walker (FLRW) background.  The chameleon
dark-energy proposal of Ref.~\cite{Cai:2021wgv} provides a representative
field-theory realization of this possibility.  A scalar field with a runaway
Peebles--Ratra potential and a matter coupling develops
\begin{equation}
V_{\rm eff}(\varphi)=V(\varphi)+\Omega(\varphi)\hat{\rho}_{\rm m},
\end{equation}
with \(V(\varphi)\propto\varphi^{-n}\), where the chameleon mass
scale is set to unity, \(\Omega(\varphi)=\exp(\varphi)\).  The ambient matter
density therefore
changes the minimum of the effective potential.  In an overdense region the
field can be trapped at a higher effective potential energy, which behaves as a
larger local dark-energy density and raises the local expansion rate relative to
the far-field background.  This differs in sign and microphysics from the
usual underdensity explanation, where a lower local matter density produces a
larger local expansion rate.  In the toy model of Ref.~\cite{Cai:2021wgv}, an
overdensity of order \(20\%\) can reproduce a local distance-ladder value of
\(H_0\) for representative parameters.  The more realistic calculation maps
BOSS Data Release 12 (DR12) LOWZ+CMASS density profiles, with galaxy-bias corrections, into a
radial \(H_0(r)\) profile using a Lema\^itre--Tolman--Bondi (LTB)-like description.  The result depends on
the assumed matter profile, the bias model, the north--south survey difference,
and the overlap between the galaxy survey and the distance indicators being
interpreted.  Two-rung ladder and intercept-based diagnostics sharpen the same
point by asking where the local-calibration information enters the distance
network~\cite{Kenworthy:2022jdh,Huang:2024gfw,Wang:2026kor}.

The chameleon construction provides a concrete local-inhomogeneity model and
exposes the required consistency conditions.  The scalar field mediates a fifth
force, so screening and laboratory or astrophysical constraints must be checked;
the local profile must relax to large-scale homogeneity; and different probes
need not sample the same regions.  Cepheid-calibrated SNe, standard
sirens, surface-brightness fluctuations, and time-delay lenses may therefore
sample different effective local expansions.  The broader local-structure
literature tests the same requirement in a less model-specific way.  It asks
whether the observed density field, possible departures from an exactly
homogeneous FLRW description, or anisotropic low-redshift expansion can generate
a sufficient \(H_0\) shift~\cite{Keenan:2013mfa,Romano:2016utn,
Shanks:2018rka,Ding:2019mmw,Haslbauer:2020xaa,Kazantzidis:2020tko,
Krishnan:2021dyb,Colin:2019opb,Yu:2022wvg}.  The common conclusion from SNe,
peculiar-velocity, BAO, CMB-dipole, and direct-distance tests is that the local
profile allowed by current data is too small or too geometrically constrained to
account for the full discrepancy without additional physics~\cite{Kenworthy:2019qwq,
 Castello:2021uad,Camarena:2022iae,Huterer:2023ldv,Banik:2025dlo,
 Mazurenko:2024gwj,Stiskalek:2025ibp,Cai:2020tpy,Moretti:2026dpo,
 Moretti:2025gbp}.  These works therefore provide
consistency tests for local-inhomogeneity mechanisms as a class.  Local voids
are unlikely to explain the full tension on their own, but environment-dependent
physics can still contribute to residual calibration or local-expansion effects.

The resulting picture is that late-time physics has direct access to the redshift
range where the discrepant measurements are made, while the acoustic ruler
remains fixed by earlier physics.  Smooth dark energy changes the whole recent
distance-redshift relation and is strongly restricted by BAO and SNe.
Localized transitions can move the relevant distance range more efficiently but
leave correlated low-redshift signatures.  Dark-sector energy transfer changes
both background densities and perturbations, making growth and lensing central
tests.  Local-inhomogeneity explanations change the meaning of the measured
local expansion rate and must be supported by a plausible matter profile and
screened microphysics.  Late-time models are therefore best regarded as partial
explanations or as components of a broader cosmological picture.  Models whose
main change is a modification of the gravitational action, field equations, or
sector-dependent gravitational response are discussed in
Sec.~\ref{sec:modified_gravity}.

\section{Outlook}\label{sec:conclusion}

This review has organized theoretical responses to the Hubble tension by the
part of the cosmological inference chain that they modify.  The central
question is how the CMB acoustic angle, the pre-recombination ruler, BAO
distances, SN relative distances, local absolute calibration, CMB spectra,
lensing, structure growth, local tests, and stability can be made mutually
consistent.  The main text therefore
first reviewed scalar-tensor gravity, non-minimally coupled quintessence, and
other modified-gravity models, then used EDE as the main representative
early-time model and discussed late-time distance, energy-transfer, and
local-inhomogeneity explanations.  The EDE chains presented here are reference calculations for
fixed-\(n\), free-\(n\), DESI-era, and DES data combinations.  They are
intended as benchmark constraints on parameter directions, not as evidence that
the data favor EDE.

\begin{table*}[t]
\caption{Summary of the approaches reviewed in this article, listing the
physical change, main observables, and leading consistency requirements.}
\label{tab:mechanism_summary}
\centering
\footnotesize
\setlength{\tabcolsep}{1pt}
\newcommand{\mechanismcell}[2]{\parbox[c]{#1}{\raggedright #2}}
\begin{tabular}{llll}
\hline\hline
\noalign{\vskip 0.6pt}
\mechanismcell{0.17\textwidth}{Approach} &
\mechanismcell{0.27\textwidth}{Physical effects} &
\mechanismcell{0.27\textwidth}{Main observables} &
\mechanismcell{0.27\textwidth}{Main limitations} \\
\hline
\noalign{\vskip 1.2pt}
\mechanismcell{0.17\textwidth}{EDE-like \(r_s\) reduction} &
\mechanismcell{0.27\textwidth}{Transient pre-recombination energy density
lowers \(r_s\) and \(r_d\)} &
\mechanismcell{0.27\textwidth}{\(H_0\), \(\omega_{\rm cdm}\), \(n_s\), damping,
CMB lensing, BAO, growth} &
\mechanismcell{0.27\textwidth}{Timing and tuning; CMB spectra, lensing, BAO, LSS,
and \(S_8\) constraints} \\
\noalign{\vskip 5pt}
\mechanismcell{0.17\textwidth}{Other pre-recombination mechanisms} &
\mechanismcell{0.27\textwidth}{Radiation, neutrino-sector, recombination,
visibility-function, transition, or early-gravity changes} &
\mechanismcell{0.27\textwidth}{CMB phase shifts,
damping tail, BBN, \(N_{\rm eff}\), \(P(k)\), late-time structure} &
\mechanismcell{0.27\textwidth}{Tight
constraints from CMB acoustic structure, BBN, perturbations, and growth} \\
\noalign{\vskip 5pt}
\mechanismcell{0.17\textwidth}{Late-time expansion dynamics} &
\mechanismcell{0.27\textwidth}{Low-redshift \(H(z)\), distance ratios, or
dark-sector energy balance after \(r_d\) is fixed} &
\mechanismcell{0.27\textwidth}{DESI BAO, SNe, inverse
distance ladder, growth, lensing, direct calibrators} &
\mechanismcell{0.27\textwidth}{Limited \(H_0\) shift
for unchanged \(r_d\); often calibrator-dependent} \\
\noalign{\vskip 5pt}
\mechanismcell{0.17\textwidth}{Local inhomogeneity} &
\mechanismcell{0.27\textwidth}{Relation between local distance
indicators and the asymptotic FLRW background} &
\mechanismcell{0.27\textwidth}{Galaxy density field, peculiar
velocities, SNe isotropy, standard sirens, time-delay lenses, CMB dipole} &
\mechanismcell{0.27\textwidth}{Matter-profile or screened-scalar requirements;
tight local-structure bounds} \\
\noalign{\vskip 5pt}
\mechanismcell{0.17\textwidth}{Modified gravity and gravity-related couplings} &
\mechanismcell{0.27\textwidth}{Gravitational action, effective Planck mass,
sector-dependent response, lensing, or growth} &
\mechanismcell{0.27\textwidth}{Background distances, CMB and weak lensing, RSD,
gravitational waves, local tests} &
\mechanismcell{0.27\textwidth}{Stability, screening, gravitational-wave speed,
local-gravity, and growth tests} \\
\noalign{\vskip 5pt}
\hline\hline
\end{tabular}
\end{table*}

Tab.~\ref{tab:mechanism_summary} also makes clear why the different routes
should be compared horizontally.  Early-time models can reduce \(r_s\) and
\(r_d\) before the standard ruler is propagated to low redshift, but the same
change tends to move \(\omega_{\rm cdm}\), \(n_s\), damping, lensing, and growth;
the timing of the transient component and the prior volume near the
\(\Lambda\)CDM boundary remain important model-building and statistical issues
~\cite{Hill:2020osr,Ivanov:2020ril,DAmico:2020ods,Poulin:2025nfb,Jedamzik:2020zmd,Pedrotti:2024kpn}.  Late-time
models act close to the redshifts at which the discrepant measurements are
made, but BAO, SNe, and inverse-distance-ladder analyses restrict the allowed
distortion of the distance-redshift relation if \(r_d\) is kept close to its
\(\Lambda\)CDM value~\cite{Bernal:2016gxb,Aylor:2018drw,Knox:2019rjx,
Schoneberg:2022ggi,Poulin:2024ken,Keeley:2022ojz,Kanodia:2025jqh}.  Modified gravity, scalar-tensor models, and
gravity-related dark-sector couplings can change both the background and
perturbation sectors, which is valuable for connecting \(H_0\) to growth or
lensing, but this strength brings stability, screening, gravitational-wave,
local-gravity, and sector-dependence requirements
~\cite{Clifton:2011jh,Braglia:2020iik,Ye:2024zpk,Pitrou:2023swx,
Uzan:2023dsk,Wang:2025znm}.  Cross-mechanism models,
including coupled or triggered EDE, neutrino-assisted transitions, dark-sector
forces, and scalar-tensor EDE-like gravity, are useful when the same physical
interaction explains more than one correlated observable
~\cite{Sakstein:2019fmf,Lin:2022phm,Trodden:2022zye,
CarrilloGonzalez:2023lma,Pitrou:2023swx,Uzan:2023dsk,Liu:2023kce,
Buen-Abad:2024tlb,Wang:2026wrk}.

Purely parameterized reconstructions play a different role.  They can identify
the shape of the expansion history required by the data without committing to a
microphysical model.  PAge is a representative example.  It parameterizes the
late-time expansion history through the cosmic age and a small number of
background parameters, giving an accurate approximation to many beyond-\(\Lambda\)CDM
expansion histories~\cite{Huang:2020mub}.  MAPAge improves this age-based
description by providing a more accurate compact reconstruction
tool~\cite{Huang:2021aku}.  PAge-style inverse-distance-ladder analyses can
therefore measure late-time \(H_0\) and curvature in a comparatively
model-independent way~\cite{Du:2025csv}.  Their interpretation is diagnostic:
they locate the required \(H(z)\) deformation, while a physical explanation
still requires a dynamical origin, a perturbation prescription, and stability
conditions.

Future progress will depend on multi-probe consistency, with a single shifted
value of \(H_0\) providing an insufficient target.  DESI BAO and full-shape
measurements, independent CMB
 spectra and lensing from ACT and SPT, future CMB polarization data, weak
 lensing, RSD, time-delay lenses, standard sirens~\cite{Giare:2024syw},
 standardized quasars~\cite{Dainotti:2024aha}, large-angle CMB/ISW
 consistency~\cite{Das:2013sca}, and improved local calibrations will test
 whether the discrepancy is tied to the
sound horizon, late expansion, local inhomogeneities, gravitational dynamics,
standard-ruler physics, or local-calibration systematics.  Theory comparisons should use common likelihoods,
consistent priors, Boltzmann-level perturbations, and both Bayesian and
profile-likelihood diagnostics.  The possible connection between axion-like EDE
and isotropic cosmic birefringence, through a rolling pseudoscalar coupled to
photons, illustrates the value of cross-probe predictions
~\cite{Murai:2022zur,Kochappan:2024jyf,Yin:2026gss,Zhang:2026fzj}.  Such
connections are best viewed as additional tests of the same microphysics and
should be separated from claims that EDE is observationally selected.

The conservative conclusion is that no theoretical explanation has yet satisfied
the full set of consistency requirements.  A successful framework must explain
\(H_0\) together with \(r_s\), \(r_d\), standard candles, CMB spectra, lensing,
structure growth, local tests, and theoretical stability.  It must also specify whether the
required shift appears without a direct local distance-ladder prior or only
after adding one.  The Hubble tension therefore remains an open problem for
precision cosmology, and its eventual explanation will have to account for the
full inference network that gives the measured value of \(H_0\) its meaning.

\appendix

\section{EDE Reference-Calculation Methodology and Data}\label{app:ede_methodology}

The original EDE and auxiliary
\(\Lambda\)CDM chains and their corresponding input files are publicly
available\footnote{\githubchain}. 
The EDE constraints shown in Sec.~\ref{sec:early} and
Appendix~\ref{app:ede_constraints} were computed with \textsc{Cobaya}\footnote{\githubcobaya}
~\cite{Torrado:2020dgo}.
The axion-like EDE chains use the \textsc{AxiCLASS}\footnote{\githubaxiclass}
implementation of the Cosmic Linear Anisotropy Solving System (\textsc{CLASS})
~\cite{Blas:2011rf,Poulin:2018cxd,Smith:2019ihp}.  In the
fixed-potential runs we set \(n=3\) and sample \(f_{\rm EDE}\),
\(\log_{10}z_c\), and \(\theta_i\), while the free-\(n\) runs sample the
 exponent as an additional parameter.  The posterior samples were obtained with
 \textsc{PolyChord}\footnote{\githubpolychord}~\cite{Handley:2015fda,Handley:2015vkr}.  Auxiliary \(\Lambda\)CDM
reference chains used for the \(H_0r_d\) diagnostic figure were run with
Code for Anisotropies in the Microwave Background(\textsc{CAMB})\footnote{\githubcamb}~\cite{Lewis:1999bs}.

The data labels follow the likelihood combinations specified in the YAML
inputs.  Planck
denotes the Planck low-\(\ell\) TT and EE likelihoods~\cite{Planck:2018vyg},
the Planck Release 4 (PR4)/NPIPE \textsc{CamSpec} high-\(\ell\) TTTEEE likelihood
~\cite{Rosenberg:2022sdy}, and Planck PR4 lensing~\cite{Carron:2022eyg}.  CMB
denotes this Planck block plus the ACT DR6 CMB-lensing likelihood with the
ACT--Planck baseline reconstruction~\cite{ACT:2023kun}.  DESI denotes the DESI
DR2 BAO measurements, based on the first three years of DESI observations
~\cite{DESI:2025zgx}.  BOSS/eBOSS denotes the 6dF Galaxy Survey, SDSS Main Galaxy Sample (MGS), and SDSS Data Release 16 (DR16) BAO
compilation used in the corresponding chains
~\cite{Beutler:2011hx,Ross:2014qpa,eBOSS:2020yzd}.  PP denotes the
PantheonPlus SN sample without SH0ES calibration
~\cite{Scolnic:2021amr,Brout:2022vxf,Riess:2021jrx}, whereas
\(+\mathrm{SH0ES}\) denotes the corresponding PantheonPlus+SH0ES
distance-ladder calibration.  DES denotes the DES-Dovekie SN
sample~\cite{DES:2025sig}, and \(+H_0\) denotes the direct local \(H_0\)
measurement added to that combination~\cite{Riess:2020fzl}.  These settings are
used as reference calculations for comparing EDE parameter directions across
recent CMB, BAO, and SN data combinations.

\section{Additional EDE Constraint Figures and Summary Table}
\label{app:ede_constraints}\label{app:ede_summary_table}

This appendix collects the EDE reference calculations that support
Sec.~\ref{sec:early} and are not included in the main discussion.  The main text
shows the fixed-\(n\) BAO comparison, the DES-Dovekie comparison, and the
free-\(n\) potential-shape test.  The table and figures below provide the full
numerical summary and three diagnostic projections.  They show the response to
local distance-ladder information, the comparison between fixed-\(n\) and
free-\(n\) chains, and the relation between model parameters and the
\(H_0\)--\(r_d\) standard-ruler plane.

Tab.~\ref{tab:ede_constraints_summary} collects the numerical EDE constraints
used in Sec.~\ref{sec:early}.  It is a compact reference for the likelihood
combinations shown in the figures.  The quoted intervals are marginalized
68\% credible intervals from the weighted chains.  For the two free-\(n\)
chains, GetDist gives \(n=3.26^{+0.75}_{-1.26}\) for CMB+DESI DR2+PP and
 \(n=3.10^{+0.26}_{-0.85}\) for CMB+DESI DR2+PP+\(\mathrm{SH0ES}\), both at
 68\% credibility.  The table reports marginalized EDE parameter
 constraints together with the best-fit \(\chi^2\) difference relative to the
 paired \(\Lambda\)CDM chain.

 \begin{table*}[t]
 \centering
 \scriptsize
 \setlength{\tabcolsep}{1.2pt}
 \caption{
 EDE constraints for all likelihood combinations used in this work.
 The entries give marginalized means with 68\% credible intervals.
 The final column gives the best-fit difference
 \(\Delta\chi^2\equiv \chi^2_{\rm EDE}
 -\chi^2_{\Lambda{\rm CDM}}\).  Negative values indicate a better EDE fit.
 These values are fit diagnostics rather than calibrated likelihood-ratio
 significances.
 Rows without an explicit free-$n$ label use the fixed benchmark value $n=3$.
 }
 \label{tab:ede_constraints_summary}
 \resizebox{\textwidth}{!}{%
 \begin{tabular}{@{}l*{9}{c}@{}}
 \hline\hline
 \noalign{\vskip 0.6pt}
 Datasets & $\Omega_{\rm b} h^2$ & $\Omega_{\rm c} h^2$ & $H_0$ & $f_{\rm EDE}$ & $\log_{10}z_c$ & $\theta_i$ & $\Omega_{\rm m}$ & $S_8$ & $\Delta\chi^2$ \\
 \hline
 \noalign{\vskip 1.2pt}
 Planck & $0.02227^{+0.00016}_{-0.00019}$ & $0.1226^{+0.0017}_{-0.0030}$ & $68.14^{+0.62}_{-1.05}$ & $0.028^{+0.007}_{-0.028}$ & $3.558^{+0.169}_{-0.229}$ & $1.92^{+1.18}_{-1.82}$ & $0.313^{+0.007}_{-0.007}$ & $0.832^{+0.011}_{-0.011}$ & $-1.2$ \\
 Planck+PP & $0.02222^{+0.00015}_{-0.00017}$ & $0.1227^{+0.0017}_{-0.0027}$ & $67.92^{+0.55}_{-0.92}$ & $0.027^{+0.007}_{-0.027}$ & $3.562^{+0.207}_{-0.225}$ & $2.12^{+0.98}_{-0.33}$ & $0.316^{+0.006}_{-0.006}$ & $0.834^{+0.011}_{-0.012}$ & $-1.0$ \\
 Planck+PP+SH0ES & $0.02257^{+0.00018}_{-0.00018}$ & $0.1307^{+0.0031}_{-0.0031}$ & $71.71^{+0.86}_{-0.97}$ & $0.119^{+0.025}_{-0.022}$ & $3.612^{+0.066}_{-0.105}$ & $2.75^{+0.18}_{-0.00}$ & $0.299^{+0.006}_{-0.006}$ & $0.835^{+0.012}_{-0.012}$ & $-26.5$ \\
 CMB & $0.02228^{+0.00016}_{-0.00018}$ & $0.1223^{+0.0016}_{-0.0026}$ & $68.09^{+0.62}_{-0.95}$ & $0.027^{+0.007}_{-0.027}$ & $3.566^{+0.243}_{-0.224}$ & $2.07^{+1.03}_{-0.32}$ & $0.313^{+0.006}_{-0.006}$ & $0.833^{+0.010}_{-0.010}$ & $-0.4$ \\
 CMB+PP & $0.02223^{+0.00015}_{-0.00017}$ & $0.1223^{+0.0013}_{-0.0025}$ & $67.76^{+0.54}_{-0.84}$ & $0.022^{+0.005}_{-0.022}$ & $3.559^{+0.266}_{-0.230}$ & $1.89^{+1.21}_{-0.48}$ & $0.316^{+0.006}_{-0.006}$ & $0.836^{+0.009}_{-0.009}$ & $-0.8$ \\
 CMB+PP+SH0ES & $0.02260^{+0.00019}_{-0.00019}$ & $0.1299^{+0.0023}_{-0.0025}$ & $71.45^{+0.80}_{-0.81}$ & $0.111^{+0.021}_{-0.019}$ & $3.595^{+0.058}_{-0.128}$ & $2.66^{+0.29}_{-0.00}$ & $0.300^{+0.005}_{-0.005}$ & $0.836^{+0.009}_{-0.009}$ & $-26.7$ \\
 CMB+eBOSS+PP & $0.02232^{+0.00015}_{-0.00017}$ & $0.1225^{+0.0016}_{-0.0029}$ & $68.39^{+0.53}_{-0.91}$ & $0.031^{+0.007}_{-0.031}$ & $3.590^{+0.269}_{-0.225}$ & $2.14^{+0.96}_{-0.28}$ & $0.311^{+0.005}_{-0.005}$ & $0.832^{+0.009}_{-0.009}$ & $-2.5$ \\
 CMB+eBOSS+PP+SH0ES & $0.02260^{+0.00021}_{-0.00022}$ & $0.1299^{+0.0026}_{-0.0029}$ & $71.23^{+0.74}_{-0.76}$ & $0.109^{+0.023}_{-0.023}$ & $3.622^{+0.241}_{-0.175}$ & $2.66^{+0.33}_{-0.00}$ & $0.302^{+0.004}_{-0.004}$ & $0.839^{+0.009}_{-0.009}$ & $-27.5$ \\
 CMB+DESI DR2+PP & $0.02243^{+0.00016}_{-0.00018}$ & $0.1222^{+0.0023}_{-0.0035}$ & $69.36^{+0.69}_{-1.08}$ & $0.044^{+0.013}_{-0.044}$ & $3.614^{+0.170}_{-0.192}$ & $2.37^{+0.73}_{-0.02}$ & $0.302^{+0.004}_{-0.004}$ & $0.822^{+0.008}_{-0.008}$ & $-3.8$ \\
 CMB+DESI DR2+PP+SH0ES & $0.02263^{+0.00018}_{-0.00019}$ & $0.1296^{+0.0026}_{-0.0027}$ & $71.75^{+0.75}_{-0.76}$ & $0.113^{+0.022}_{-0.021}$ & $3.610^{+0.070}_{-0.123}$ & $2.71^{+0.24}_{-0.00}$ & $0.297^{+0.003}_{-0.003}$ & $0.832^{+0.009}_{-0.008}$ & $-28.3$ \\
 CMB+DES & $0.02223^{+0.00015}_{-0.00017}$ & $0.1226^{+0.0012}_{-0.0026}$ & $67.78^{+0.51}_{-0.92}$ & $0.023^{+0.004}_{-0.023}$ & $3.536^{+0.193}_{-0.217}$ & $1.88^{+1.22}_{-0.49}$ & $0.317^{+0.006}_{-0.006}$ & $0.838^{+0.010}_{-0.010}$ & $0.0$ \\
 CMB+DES+$H_0$ & $0.02247^{+0.00018}_{-0.00018}$ & $0.1264^{+0.0027}_{-0.0027}$ & $69.64^{+0.77}_{-0.80}$ & $0.070^{+0.026}_{-0.021}$ & $3.611^{+0.249}_{-0.202}$ & $2.45^{+0.59}_{-0.01}$ & $0.308^{+0.005}_{-0.005}$ & $0.837^{+0.009}_{-0.009}$ & $-11.5$ \\
 \multicolumn{10}{l}{\textbf{free \(n\)}} \\
 CMB+DESI DR2+PP & $0.02242^{+0.00016}_{-0.00019}$ & $0.1228^{+0.0023}_{-0.0039}$ & $69.51^{+0.73}_{-1.11}$ & $0.047^{+0.016}_{-0.044}$ & $3.617^{+0.335}_{-0.242}$ & $2.35^{+0.75}_{-0.01}$ & $0.302^{+0.003}_{-0.004}$ & $0.822^{+0.009}_{-0.009}$ & $-4.2$ \\
 CMB+DESI DR2+PP+SH0ES & $0.02256^{+0.00022}_{-0.00024}$ & $0.1300^{+0.0026}_{-0.0034}$ & $71.78^{+0.72}_{-0.79}$ & $0.114^{+0.024}_{-0.023}$ & $3.590^{+0.056}_{-0.119}$ & $2.69^{+0.30}_{-0.10}$ & $0.297^{+0.003}_{-0.003}$ & $0.832^{+0.008}_{-0.009}$ & $-28.3$ \\
\hline\hline
\end{tabular}%
}
\end{table*}

The best-fit values show a clear contrast between combinations with
and without local distance-ladder information.  Without SH0ES or the direct
local \(H_0\) likelihood, the reduction in the best-fit \(\chi^2\) does not
exceed 4.2.  The PantheonPlus+SH0ES combinations yield \(\Delta\chi^2\) values
between \(-28.3\) and \(-26.5\), while CMB+DES+\(H_0\) gives \(-11.5\).
Within the combinations examined here, the large best-fit improvements
therefore arise only after adding SH0ES calibration or a direct local \(H_0\)
likelihood.

Several qualitative patterns are visible in the table.  Planck-only,
CMB-only, and uncalibrated PantheonPlus combinations keep \(f_{\rm EDE}\) near
the low boundary and prefer lower \(H_0\).  Among the uncalibrated PP
combinations, DESI DR2 permits a broader extension toward higher
\(f_{\rm EDE}\) and \(H_0\) than the PP-only and BOSS/eBOSS cases, but does not
select the SH0ES-like high-\(H_0\) region.  Adding SH0ES calibration moves the
PP-based posteriors to \(f_{\rm EDE}\simeq0.11\) and
\(H_0\simeq71\)--\(72~{\rm km\,s^{-1}\,Mpc^{-1}}\).  The DES combination with
a direct local \(H_0\) measurement lies between the uncalibrated and
SH0ES-calibrated PP results.  The free-\(n\) rows show the same strong
dependence on whether SH0ES calibration is included.

\begin{figure}[!htbp]
\centering
\includegraphics[width=\linewidth,keepaspectratio]{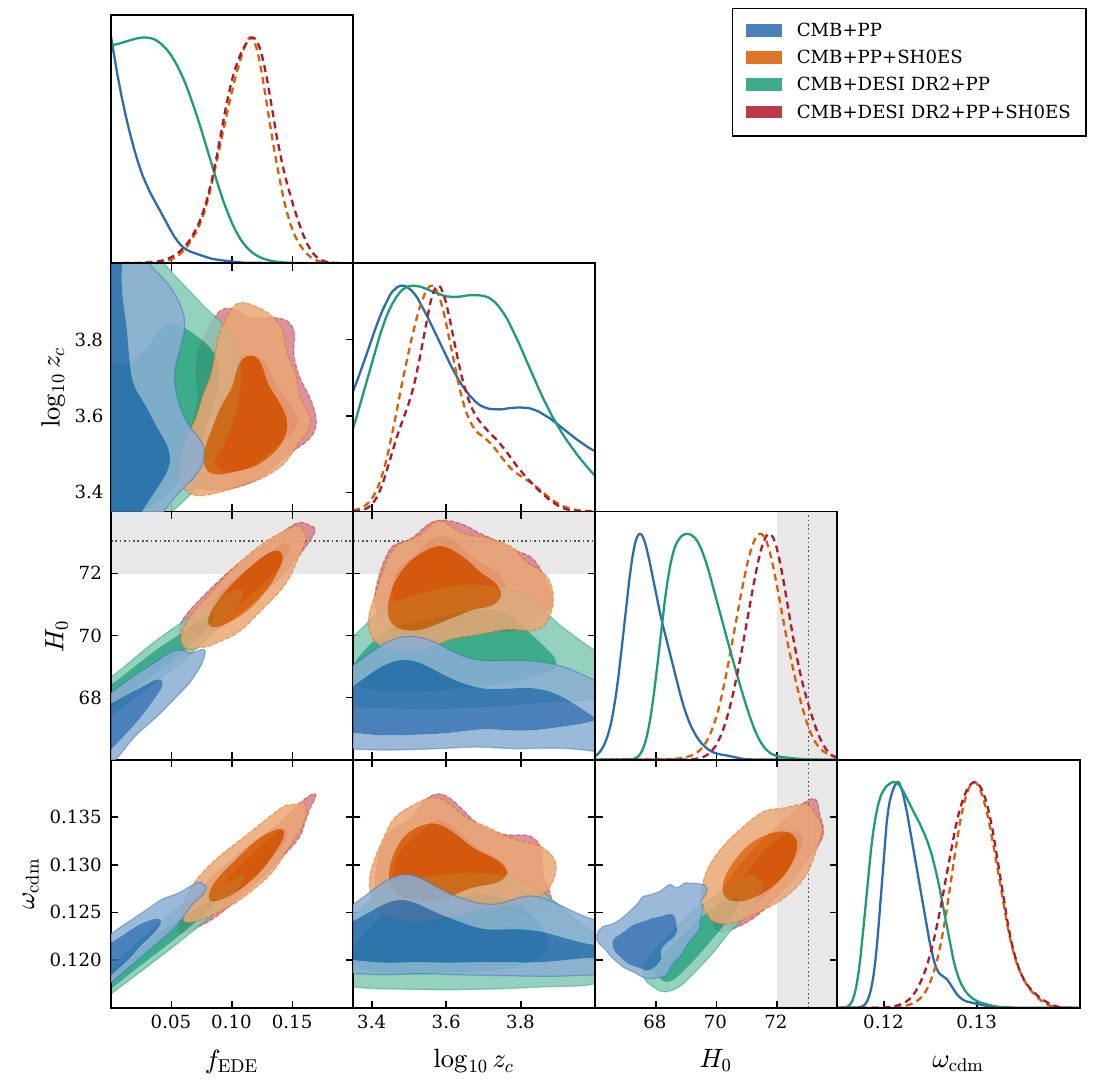}
\caption{
Fixed-$n$ EDE reference chains with and without SH0ES calibration.  Solid
and dashed contours denote data combinations without and with SH0ES,
respectively.
}
\label{fig:ede_h0_prior_appendix}
\end{figure}

The first diagnostic focuses on local distance-ladder information in the
PantheonPlus and DESI DR2 combinations.  Fig.~\ref{fig:ede_h0_prior_appendix}
compares fixed-$n$ chains with and without SH0ES calibration.  Adding
SH0ES produces a clear shift along the existing high-$f_{\rm EDE}$,
high-$H_0$, high-$\omega_{\rm cdm}$ direction.  The CMB, BAO, and SN
likelihoods continue to determine the allowed width and limit one-parameter
shifts of $H_0$.

\begin{figure}[!htbp]
\centering
\includegraphics[width=\linewidth,keepaspectratio]{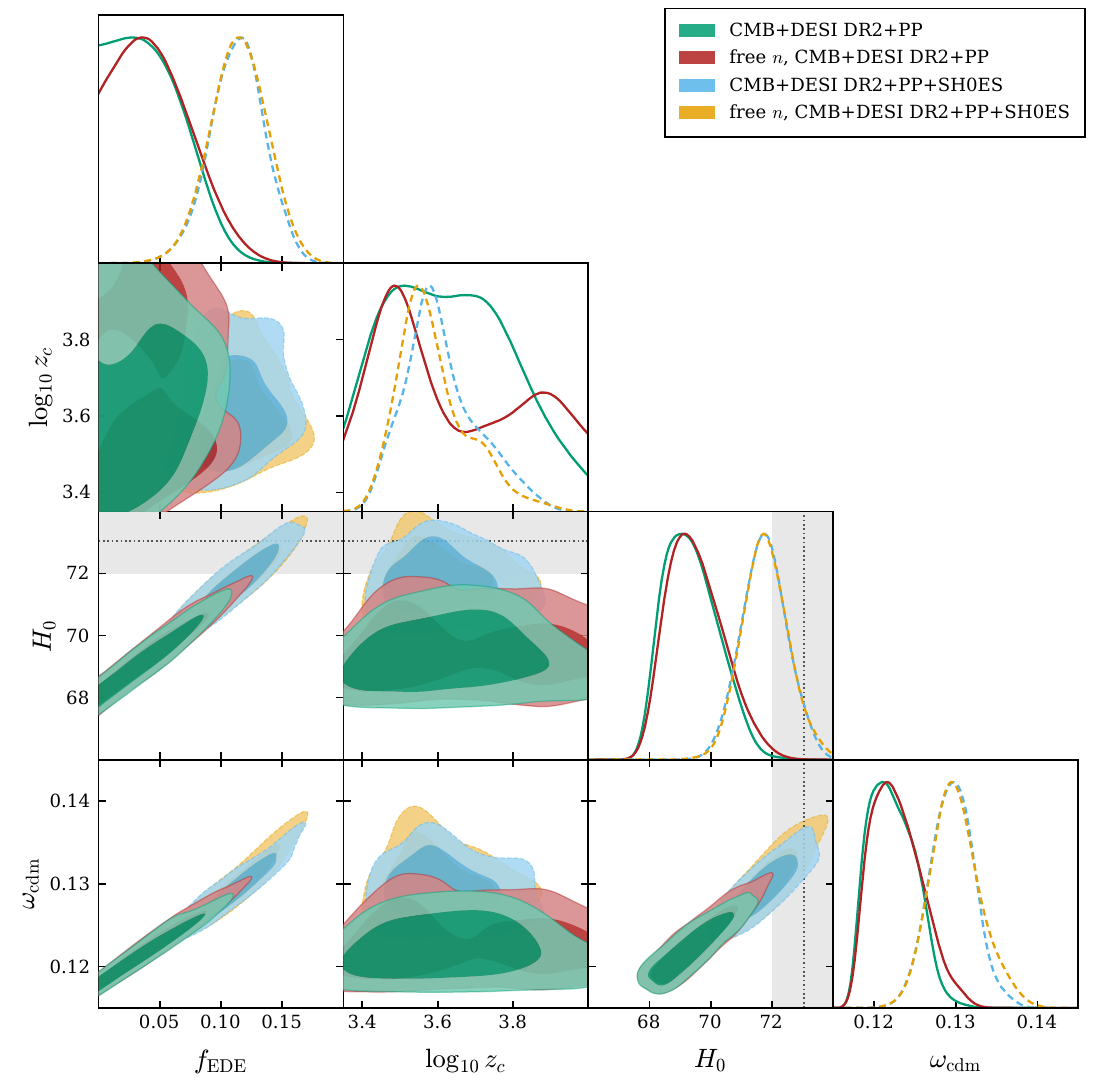}
\caption{
Comparison of fixed-$n$ and free-$n$ EDE constraints on the common parameter
space.  The plotted parameters are \(f_{\rm EDE}\), \(\log_{10}z_c\), \(H_0\),
and \(\omega_{\rm cdm}\).
}
\label{fig:ede_fixedn_vs_freen_appendix}
\end{figure}

The second diagnostic compares the fixed-$n$ and free-$n$ chains in a common
parameter projection.  Fig.~\ref{fig:ede_fixedn_vs_freen_appendix} shows that
allowing the exponent to vary changes the EDE posterior, especially in the
transition-redshift direction.  The common projection still preserves the
correlated $H_0$--$\omega_{\rm cdm}$ response that controls the standard
axion-like EDE fit.

\begin{figure}[!htbp]
\centering
\includegraphics[width=\linewidth,keepaspectratio]{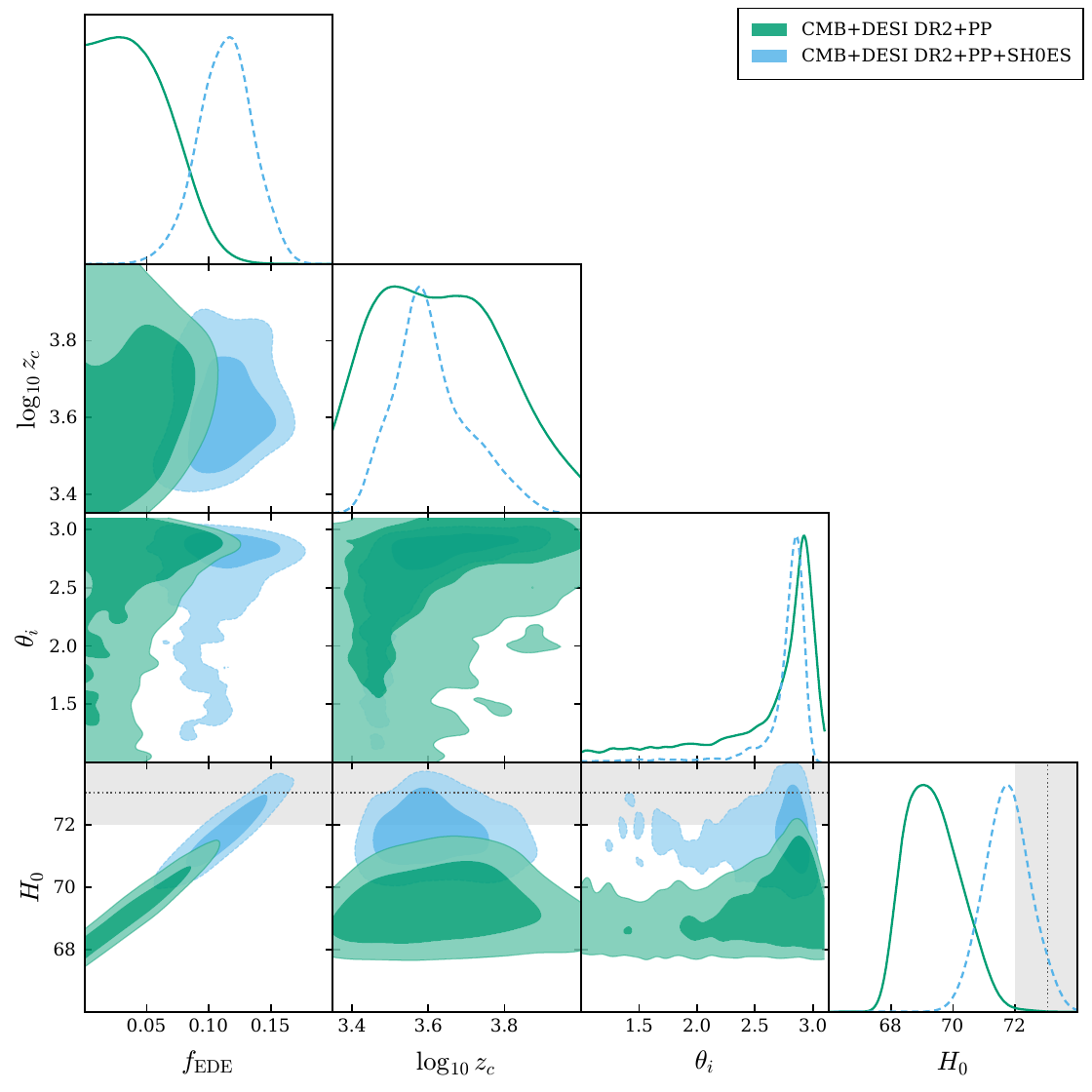}
\caption{
Model-parameter triangle for fixed-$n$ EDE, including the initial displacement
\(\theta_i\).  The plotted parameters are \(f_{\rm EDE}\), \(\log_{10}z_c\),
\(\theta_i\), and \(H_0\).
}
\label{fig:ede_model_parameter_appendix}
\end{figure}

The third diagnostic displays the internal model-parameter structure of the
fixed-$n$ chains.  Fig.~\ref{fig:ede_model_parameter_appendix} includes
$\theta_i$, which is omitted from the main-text figure to keep the physical
comparison readable.  The concentration of posterior weight near a large
initial displacement is part of the model-building cost discussed in
Sec.~\ref{sec:early}.  It expresses the requirement that the field reach an
appreciable fractional density near the epoch where a reduction of the sound
horizon is effective.

\begin{figure}[!htbp]
\centering
\includegraphics[width=\linewidth,keepaspectratio]{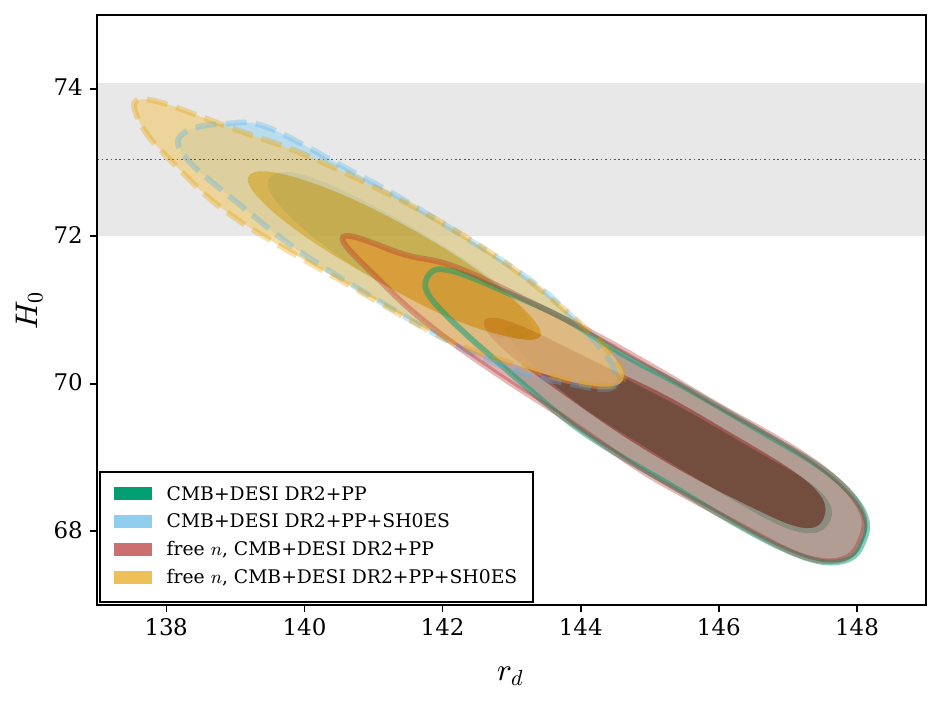}
\caption{
\(H_0\)--\(r_d\) plane for the EDE reference chains.  The gray band marks the
local distance-ladder reference range used consistently in these EDE figures.
}
\label{fig:ede_h0_rd_appendix}
\end{figure}

Finally, Fig.~\ref{fig:ede_h0_rd_appendix} displays the same reference
calculations in the $H_0$--$r_d$ plane.  This projection shows the
standard-ruler requirement directly.  A higher inferred $H_0$ in an early-time
standard-ruler scenario is accompanied by a smaller drag-epoch ruler.  The
remaining question,
addressed by the CMB spectra, BAO distances, SN likelihoods, and
growth-sensitive data, is whether that smaller ruler can be embedded in a
consistent cosmological fit.

\begin{acknowledgements}
This work is supported in part by the National Natural Science Foundation of China Grants No. 12475067 and No. 12235019.
We also acknowledge the use of the HPC Cluster of ITP-CAS.
\end{acknowledgements}

\bibliographystyle{raa}
\bibliography{refer}

\end{document}